\documentclass{article}

\usepackage[utf8]{inputenc}
\usepackage[T1]{fontenc}

\usepackage{setspace}
\usepackage{amssymb}
\usepackage{amsmath,amsthm}
\usepackage{sansmath}
\usepackage{mathtools}
\usepackage{authblk}
\usepackage[dvipdfmx]{graphicx} 
\usepackage{bmpsize}
\usepackage[hyphens]{url}

\usepackage{enumitem}
\usepackage{algorithm}
\usepackage{booktabs}
\usepackage{bm}
\usepackage[toc,page]{appendix}		
\usepackage{subcaption} 			
\usepackage{array}

\usepackage{textcomp}

\usepackage{pdfpages}
\usepackage{algpseudocode}
\usepackage[margin=0.5in]{caption}
\usepackage{comment}
\usepackage{csquotes}
\usepackage{threeparttable}
\usepackage[style=numeric,uniquename=false,uniquelist=false,date=year,backend=biber,maxbibnames=19,useprefix=true,giveninits=true]{biblatex}
\DeclareNameAlias{sortname}{family-given} 
\renewbibmacro{in:}{} 
\DeclareDelimFormat{nameyeardelim}{\addcomma\space} 
\usepackage[margin=1in,headheight=14pt]{geometry}
\usepackage{fancyhdr}
\usepackage[export]{adjustbox} 
\usepackage[section]{placeins}

\usepackage{algorithmicx}
\usepackage{algorithm}
\usepackage{etoolbox} 

\usepackage[breaklinks=true,colorlinks=true,linkcolor=black,anchorcolor=black,citecolor=black,filecolor=black,menucolor=black,runcolor=black,urlcolor=black]{hyperref}

\newcommand{\Beta}{\mathrm{B}}

\makeatletter
\newenvironment{breakablealgorithm}
{
	\begin{center}
		\refstepcounter{algorithm}
		\hrule height.8pt depth0pt \kern2pt
		\renewcommand{\caption}[2][\relax]{
			{\raggedright\textbf{\fname@algorithm~\thealgorithm} ##2\par}%
			\ifx\relax##1\relax 
			\addcontentsline{loa}{algorithm}{\protect\numberline{\thealgorithm}##2}%
			\else 
			\addcontentsline{loa}{algorithm}{\protect\numberline{\thealgorithm}##1}%
			\fi
			\kern2pt\hrule\kern2pt
		}
	}{
		\kern2pt\hrule\relax
	\end{center}
}
\makeatother

\begin{document}
\title{Bayesian multilevel multivariate logistic regression for superiority decision-making under observable treatment heterogeneity}
\date{ }
\author[1,3]{X.M. Kavelaars\thanks{E-mail: \texttt{x.m.kavelaars@tilburguniversity.edu}}}
\author[1]{J. Mulder}
\author[2]{M.C. Kaptein} 

\affil[1]{Department of Methodology and Statistics, Tilburg University, Tilburg, The Netherlands}
\affil[2]{Jheronimus Academy of Data Science, 's Hertogenbosch, The Netherlands}
\affil[3]{Department of Theory, Methodology, and Statistics, Open University of the Netherlands, Heerlen, The Netherlands}
\date{ }
	
\maketitle 

\newpage
	\begin{abstract}
\textbf{Background:} 
In medical, social, and behavioral research we often encounter datasets with a multilevel structure and multiple correlated dependent variables.
These data are frequently collected from a study population that distinguishes several subpopulations with different (i.e., heterogeneous) effects of an intervention. 	
Despite the frequent occurrence of such data, methods to analyze them are less common and researchers often resort to either ignoring the multilevel and/or heterogeneous structure, analyzing only a single dependent variable, or a combination of these.
These analysis strategies are suboptimal: Ignoring multilevel structures inflates Type I error rates, while neglecting the multivariate or heterogeneous structure masks detailed insights.

\textbf{Methods:} 
To analyze such data comprehensively, the current paper presents a novel Bayesian multilevel multivariate logistic regression model.
The clustered structure of multilevel data is taken into account, such that posterior inferences can be made with accurate error rates. 	
Further, the model shares information between different subpopulations in the estimation of average and conditional average multivariate treatment effects.
To facilitate interpretation, multivariate logistic regression parameters are transformed to posterior success probabilities and differences between them.

\textbf{Results:}
A numerical evaluation compared our framework to less comprehensive alternatives and highlighted the need to model the multilevel structure: Treatment comparisons based on the multilevel model had targeted Type I error rates, while single-level alternatives resulted in inflated Type I errors. Further, the multilevel model was more powerful than a single-level model when the number of clusters was higher.
A re-analysis of the Third International Stroke Trial data illustrated how incorporating a multilevel structure, assessing treatment heterogeneity, and combining dependent variables contributed to an in-depth understanding of treatment effects. Further, we demonstrated how Bayes factors can aid in the selection of a suitable model. 

\textbf{Conclusion:}
The method is useful in prediction of treatment effects and decision-making within subpopulations from multiple clusters, while taking advantage of the size of the entire study sample and while properly incorporating the uncertainty in a principled probabilistic manner using the full posterior distribution.

\end{abstract}

\newpage
\section{Background}

In medical, social, and behavioral research we often encounter datasets with a multilevel structure and multiple correlated dependent variables.
An example of such a study is the Cognition and Radiation Study B \parencite[][]{Schimmel2018,Schimmel2022} that investigated whether local brain radiation (stereotactic radiosurgery) preserves cognitive functioning and quality of life better than whole brain radiation in cancer patients with multiple brain metastases. 
Patients were recruited from multiple hospitals and the treatment was executed in two treatment centers, giving the data a multilevel structure. 
Many other examples of such datasets can be found in a paper by Biswas and colleagues \cite{Biswas2009}, who presented a nonexhaustive overview of hundreds of Bayesian trial protocols executed in a specialized center for cancer treatment.
The authors noted that a) almost half of the reviewed studies were multicenter trials; and b) many studies were designed to assess effectiveness and side effects simultaneously, thus including at least two dependent variables.

Often, these multilevel, multivariate data are collected from a study population that consists of several subpopulations with potentially distinctive (i.e., heterogeneous) effects of an intervention. 
Examples of such studies are the two International Stroke Trials \parencite[International Stroke Trial (IST) and Third International Stroke Trial (IST-3);][]{ISTCG1997,Sandercock2011,IST2012,Sandercock2016}, which investigated the effects of antiplatelet and antithrombotic treatments on various (neuro)psychological, functional and psychosocial dependent variables respectively. 
Both trials covered multiple treatment centers from multiple countries and included a variety of patient characteristics that could potentially predict treatment effects.
We discuss the IST-3 in more depth as it serves as a running example throughout the paper.
The IST-3 investigated the effects of an intravenous thrombolysis treatment on shortterm (e.g., recurrent stroke, functional deficits) and long-term (e.g., dependency, depression, pain) indicators of health status among patients who suffered from an acute ischaemic stroke. 
The IST-3 data revealed considerable variation in characteristics of patients and disease - such as subtype or severity of stroke, blood pressure, and age - that can be predictive of treatment effects and call for exploration of treatment heterogeneity to gain insight into subpopulation-specific effects \parencite{Lindley2015}.

All of the abovementioned trials made treatment comparisons in the context of Randomized Controlled Trials (RCTs): 
Randomized experiments in which an experimental or a control treatment is randomly assigned and administered to a random sample of patients.
RCTs often aim to evaluate whether the experimental treatment is superior or (non-)inferior to the control condition and ultimately guide clinicians in evidence-based assignment of treatments and interventions \parencite{FDA2016}.

Whereas RCTs are considered a golden standard for treatment comparison, their implementation is challenged by a growing demand for personalized treatment \parencite{Evans2003,Ng2009,Grol2003,Simon2010}. 
That is, clinical practice relies more and more on the idea that different patients react differently to treatments.
Treatment prescription is increasingly guided by a trade-off between patient-specific risks and benefits, making the research context for these decisions multivariate and heterogeneous \parencite{Murray2016}.
While demanding more complex methodology, personalization of treatments can impede the collection of sufficient data for rigorous treatment evaluation.
Development of more targeted treatments limits eligibility for participation in trials, thereby making the recruitment of subjects more difficult.
As a solution, trials more often span multiple treatment centers or countries.
This adds another layer of complexity to the research context: clustered data that require multilevel analysis. 
To meet the methodological demands of these increasingly complex research problems, RCTs ideally provide a) a broad understanding of the treatment's effects on multiple dependent variables; and b) insights potential dependencies of treatment effects on characteristics of patients; and c) an accurate handling of clustered data structures.
In practice, such comprehensive methods are less common, and often researchers resort to either ignoring the multilevel and/or heterogeneous structure, analyzing only a single dependent variable, or a combination of these.
Below, we discuss how the abovementioned three aspects can be implemented in Randomized Controlled Trial methodology to support research in personalized treatment.

First, many RCTs evaluate more than one dependent variable, which are analysed separately in multiple univariate analyses \parencite{FDA2017}.
As an example, the investigators of the IST-3 were primarily interested in living independently six months after stroke and secondarily in several other dependent variables, such as recurrent events, adverse reactions to the treatment, and mental health indicators.
Analyzing dependent variables independently provides useful insights in treatment effects on each of these dependent variables individually, but discards available information about the relation between them.
When the effects on individual dependent variables are complemented with information about their co-occurrences via multivariate analysis, a more detailed picture of treatment effects emerges.
Multivariate analysis models relationships between dependent variables and can a) be helpful to detect outcome patterns that would be ignored when dependent variables are considered in isolation; and b) improve the accuracy of sample size computations and error rates in statistical decision-making \parencite{FDA2017,Su2012,Sozu2016,Kavelaars2020}.

Second, incorporating patient and/or disease characteristics in treatment comparison can result in a considerable improvement of the practical value of RCTs. 
The IST-3 used a sample of diverse patients with different personal and disease characteristics. 
This variation contains valuable information regarding differences in treatment effects. 
For example, knowing whether patients with different weights or blood pressures have different chances of a recurrent stroke or independent living has the potential to inform treatment recommendations.
When treatments have distinct effects on patients with different characteristics, treatment effects are considered heterogeneous among (sub)populations of patients.  
In this case, average treatment effects (ATEs) give a global idea of treatment results among the trial population, but have limited value in targeting treatments to specific patients with their individual (disease) characteristics \parencite{Hamburg2010,Mirnezami2012,Schork2015}. 
Conditional average treatment effects (CATEs) among specific patient groups provide insight in the variation of treatment effects among the population and help to distinguish patients who ultimately benefit from the treatment from those who do not or may even experience adverse treatment effects.
Unfortunately, subgroup-specific treatment comparisons are insufficiently implemented as part of standard trial methodology yet \parencite{Thall2020}.
If subgroups are targeted at all, their effects are often analyzed independently via stratified (or subgroup) analysis.
Such a subgroup analysis disregards information from related subgroups and suffers from suboptimal power due to subsetting. 
Modelling heterogeneity is a more powerful alternative that directly uses the relation between subgroups and allows subgroups to borrow strength from each other \parencite{Kaptein2014,Kaptein2015,Kavelaars2022}.

Third, multilevel data are characterized by observational units that are grouped in clusters.
For example, the IST-3 spans multiple treatment centers and multiple countries.
Reasons to use multilevel analysis can be both substantive and statistical. 	
From a substantive perspective, multilevel analysis can be useful to explain differences between clusters, while using the information from the entire sample \parencite{Viele2014,Gelman2007}. 
Different trials may - for example - have overlapping but non-identical target populations that can be distinguished by covariate information and may contribute to the understanding of treatment effects.
Statistically, differences between clusters should be taken into account for the sake of validity, even if these differences are not of direct interest \parencite{Hox2017,Raudenbush2001,McGlothlin2018}.
Clustered data require specific analysis methods that are flexible enough to treat observations from different clusters as more similar to each other than to observations from other clusters.
If observations within clusters are indeed more similar, the clustered structure is reflected in variance partitioning, where the within-cluster and the between-cluster variances are modelled separately. This induces a dependence between the observations within clusters when marginalizing over the cluster-specific effects. 
When clustered observations are treated as independent observations on the other hand, variance originating from differences between clusters is then erroneously attributed to differences between a manifold of observational units and the unique amount of information is overestimated.
As a result, standard errors are overestimated, Type I error rates are inflated, and validity of statistical inference is compromised.
The larger the variance between clusters relative to the variance between observational units within clusters, the larger the effect on standard errors.
Properly modelling the multilevel structure of clustered data and allowing the parameters to vary over clusters is therefore crucial for accurate statistical decision-making \parencite{Hox2017,Raudenbush2001}.

The current paper presents a Bayesian multilevel multivariate logistic regression (BMMLR) framework to capture the three abovementioned methodological aspects in a comprehensive analysis and decision procedure for treatment comparison. 
We build upon an existing Bayesian multivariate logistic regression (BMLR) framework for single-level data to analyze multivariate binary data in the presence of treatment heterogeneity and present a multilevel extension to deal with multilevel data. 
The multilevel aspect adds another layer of complexity, making the analysis a non-trivial endeavour.
We discuss the existing BMLR framework first.
This framework consists of three coherent elements \parencite{Kavelaars2022}:
\begin{enumerate}
	\item a multivariate modelling procedure to find unknown regression parameters;
	\item a transformation procedure to convert regression parameters to the  probability scale to make analysis results more interpretable;
	\item a compatible decision procedure to draw conclusions regarding treatment superiority or inferiority with targeted Type I error rates.
\end{enumerate}  
The first element, the modelling procedure, assumes multivariate Bernoulli distributed dependent variables and assigns them a multinomial parametrization. 
A multinomial parametrization is helpful for two reasons, since it a) allows statisticians to draw and build upon existing, established multinomial techniques with tractable (conditional) posterior distributions; and b) has the flexibility to model correlations between dependent variables on the subpopulation level, which contributes to the accuracy of inference under treatment heterogeneity \parencite{Dai2013,Kavelaars2020,Kavelaars2022}.
Several other multivariate modelling procedures, such as the multivariate probit model \cite{Chib1995} or multivariate logistic regression models \cite{Malik1973, OBrien2004}, have a more restrictive correlation structure and are therefore theoretically less suitable to detect treatment heterogeneity with adequate error control.
Moreover, the multivariate logistic regression model by Malik and Abraham \cite{Malik1973} does not provide insight in the treatment effects on individual dependent variables.
Copula structures have been proposed as promising multivariate alternatives as well, but these models can be difficult to apply to binary dependent variables \parencite{Braeken2007,Nikoloulopoulos2008,Panagiotelis2012}. 
The second element, the transformation procedure, builds upon the close relation between the multinomial and multivariate parametrizations to express results on the scale of (multivariate) success probabilities and differences between them, as a more intuitive alternative to multinomial (log-)odds.
The transformed parameters provide understandable insights in the treatment's performance on the trial population (i.e., ATEs) as well as subpopulations of interest (i.e., CATEs).
The third element, the decision procedure, conveniently uses the Bayesian nature of the modelling procedure, allowing for inference on the posterior samples of transformed parameters. 
Decisions can be made in several ways to flexibly combine and weigh multiple dependent variables into a single decision for a population of interest, while taking correlations between dependent variables into account.

The main contribution of the current paper is the extension of the single-level BMLR framework to the multilevel context.
The novel Bayesian multilevel multivariate logistic regression (BMMLR) framework provides BMLR with a multilevel model component and adjusts the transformation and decision procedure accordingly, to make the framework suitable for the multilevel context, resulting in accurate type I errors. 
The remainder of the paper is structured as follows.
Section \ref{s:model} introduces the multilevel multivariate logistic regression model to obtain a sample from the posterior distribution of regression coefficients. 
Section \ref{s:treatmentcomparison} outlines how to transform the obtained regression coefficients to more interpretable treatment effect parameters.
Section \ref{s:decision} discusses the decision procedure to use the treatment effect parameters for treatment comparison.
Section \ref{s:evaluation} demonstrates the performance of the model numerically via simulation and in Section \ref{s:application} the methodology is illustrated with data from the IST-3. 
The paper concludes with a discussion in Section \ref{s:discussion}.

\section{BMMLR: Bayesian multilevel multivariate logistic regression }\label{s:model}

Consider the general case with $K \in \{1,\dots,K\}$ binary dependent variables $y^{k}_{ji}$ for subject $i \in \{1,\dots,n_{j}\}$ in cluster $j \in \{1,\dots,J\}$. 
Outcome $y^{k}_{ji}$ is Bernoulli distributed with success probability $\theta^{k}_{ji}$ and multivariate vector of $K$ dependent variables, $\bm{y}_{ji} = (y^{1}_{ji}, \dots, y^{K}_{ji})$ is multivariate Bernoulli distributed \parencite{Dai2013}. 
The multivariate Bernoulli distribution relies on a hybrid parameterization where a $K$-variate success probability in $\bm{\theta}_{ji} = (\theta^{1}_{ji}, \dots, \theta^{K}_{ji})$ is expressed in terms of $Q = 2^{K}$ multinomial joint response probabilities in $\bm{\phi}_{ji} = (\phi^{1}_{ji}, \dots, \phi^{Q}_{ji})$ \parencite{Dai2013}.
The $q^{\text{th}}$ joint response probability in $\bm{\phi}_{ji}$ corresponds to multinomial response combination $\bm{h}^{q}$, which has length $K$ and is given in the $q^{th}$ row of the matrix of joint response combinations denoted by $\bm{H}$: 
\begin{flalign}
	\bm{H} = & 
	\begin{bsmallmatrix}
		1 & 1 & \dots & 1 & 1 \\
		1 &1 & \dots & 1 & 0 \\
		& & \dots & &\\
		0 & 0 & \dots & 0 & 1 \\
		0 & 0 & \dots & 0 & 0\\
	\end{bsmallmatrix}
\end{flalign}
Hence, joint response probability $\phi^{q}_{ji} = p(\bm{y}_{ji} = \bm{h}^{q})$.
Note that the joint response probability $\bm{\phi}_{j}$ and the success probability $\bm{\theta}_{j}$ are identical in the univariate situation (i.e., $K=1$).

\subsection{Likelihood of the data}\label{ss:likelihood}

The multinomial parametrization of multivariately Bernoulli distributed data allows to model the relation between dependent variables $\bm{y}_{ji}$ and one or multiple predictor variables via multinomial logistic regression.
Joint response probability $\phi^{q}_{ji}$ is then regressed on a vector of $P$ covariates, $\bm{x}_{ji} = (x_{ji0}, \dots, x_{ji(P-1)})$.
Covariate $x_{ji0} = 1$ is a constant to estimate the intercept and covariate $x_{jip}$ for $p \in \{1,\dots,P-1\}$ can, for example, be a treatment indicator, a patient characteristic, or an interaction between these.

The relation between outcome vector $\bm{y}_{ji}$ and covariate vector $\bm{x}_{ji}$ is mapped with a multinomial logistic function that expresses the probability of $\bm{y}_{ji}$ being in response category $q$, conditional on $\bm{x}_{ji}$:
\begin{flalign}\label{eq:H_lik_mult}
	\phi^{q}_{ji} & = 
	p(\bm{y}_{ji} = \bm{h}^{q} | \bm{x}_{ji}) \\\nonumber
	& = \frac{\exp(\psi^{q}_{ji})
	}{ \displaystyle\sum_{r=1}^{Q-1} \exp(\psi^{r}_{ji})
		+ 1}, \nonumber
\end{flalign}
\noindent Here, $\psi^{q}_{ji}$
is a linear predictor:
\begin{flalign}\label{eq:NH_psi}
	\psi^{q}_{ji} 
	= & \bm{x}^{'}_{ji} \bm{\gamma}^{q}_{j}
\end{flalign}

\noindent 
In Equation \ref{eq:NH_psi}, regression coefficients for response category $q$, $\bm{\gamma}^{q}_{j} = (\gamma^{q}_{0j},\dots,\gamma^{q}_{(P-1)j})$ are unknown parameters of interest.
Regression coefficients of response categories $1,\dots,Q-1$ are estimated, while regression coefficients of response category $Q$ are fixed at zero (i.e., $\bm{\gamma}^{Q}_{j} = \bm{0}$) to ensure identifiability of the model.
The entire set of regression coefficients in cluster $j$ is denoted with $\bm{\gamma}_{j}$.

\label{R2_1_BMMLR}
A key aspect of multilevel models is that the regression coefficients $\bm{\gamma}_{j}^{q}$ are allowed to vary over clusters according to a common normal distribution on the second level. 
The common distribution for the random effects on the second level induces a dependency structure of the observations within clusters. 
The observations of diffferent individuals in the same clusters are assumed to be conditionally independent conditional on the cluster-specific random effects. 
The random effects distribution on the second level can be written as:
\begin{flalign}\label{eq:H_RC}
	\gamma^{q}_{pj} & = \gamma^{q}_{p0} + 
	u^{q}_{pj}\\\nonumber
	\bm{u}^{q}_{j}   & = (u^{q}_{0j},\dots,u^{q}_{(P-1)j}) \sim N(\bm{0},\bm{\Sigma}^{q})\nonumber
\end{flalign}
Equation \ref{eq:H_RC} consists of two elements that reflect the distributional parameters: 
\begin{enumerate}
	\item  The parameter $\gamma^{q}_{p0}$ is the common effect in the population and does not vary over clusters.
	\item The random effect $u^{q}_{pj}$ quantifies the cluster specific deviation from the common effect $\gamma^{q}_{p0}$. 
\end{enumerate}  
Equation \ref{eq:H_RC} can be adjusted to model cluster-specific predictors or cross-level interactions between cluster-level predictors and individual level-predictors. 
Further, Equation \ref{eq:H_RC} can be extended to model mixed effects, which combine regression coefficients that vary over clusters, which are called random effects, and regression coefficients that are identical for all clusters, which are called fixed effects.
More information on the specification of more complex linear predictors can be found in general resources on multilevel models, such as Hox et al. \cite{Hox2017} or Gelman and Hill \cite{Gelman2007}.
In general, it should be noted that each additional random effect increases the number of parameters, affecting computational burden and estimation precision.

\subsection{Posterior distribution of regression coefficients}\label{ss:posterior_computation}

The primary goal of BMMLR is estimating the joint posterior distribution of unknown regression coefficients $\bm{\gamma}^{q}_{j}$, their means $\bm{\gamma}^{q}$, and their covariance matrices $\bm{\Sigma}^{q}$ for category $q \in 1,\dots,(Q-1)$.
The posterior probability distribution of these parameters for category $q$ is given by:

\begin{flalign}\label{eq:post}
	p(\bm{\gamma}^{q}_{j}, \bm{\gamma}^{q}, \bm{\Sigma}^{q} | \bm{y}) \propto &
	p(\bm{y}_{j}|\bm{\gamma}^{q}_{j}) p(\bm{\gamma}^{q}_{j} | \bm{\gamma}^{q}, \bm{\Sigma}^{q}) p(\bm{\gamma}^{q}) p(\bm{\Sigma}^{q}), 
\end{flalign}%
%
%
\noindent where $\bm\gamma^{q}$ reflects the vector of average effects for category $q$, $\bm\Sigma^{q}$ is the covariance matrix of the effects across clusters for category $q$, and $\bm\gamma_{j}^{q}$ reflects the vector of cluster specific effects of cluster $j$ for category $q$.
The posterior probability distribution in Equation \ref{eq:post} is proportional to the product of three types of probability distributions: 
\begin{enumerate}
	\item The likelihood of the data quantifies the probability of the dependent variables conditional on cluster-specific regression coefficients, $p(\bm{y}_{j}|\bm{\gamma}^{q}_{j})$, which is the multinomial logistic function given by Equation \ref{eq:H_lik_mult};
	\item The probability distribution of the cluster-specific regression coefficients $\bm{\gamma}^{q}_{j}$ conditional on their means $\bm{\gamma}^{q}$ and covariance matrix $\bm{\Sigma}^{q}$ for category $q$, $p(\bm{\gamma}^{q}_{j} | \bm{\gamma}^{q}, \bm{\Sigma}^{q})$;
	\item The prior probability distributions of regression coefficient's means $\bm{\gamma}^{q}$, $p(\bm{\gamma}^{q})$, and covariance matrix $\bm{\Sigma}^{q}$, $p(\bm{\Sigma}^{q})$ for category $q$, before observing the data.
\end{enumerate}

As the multinomial logistic function (Equation \ref{eq:H_lik_mult}) does not have a (conditionally) conjugate prior distribution, the functional form of the posterior distribution is unknown and the regression coefficients cannot be sampled directly from the posterior distribution. 
In the Supplemental material, we present a Gibbs sampling algorithm based on a P\'olya-Gamma auxiliary variable expansion of the likelihood proposed by Polson et al. \cite{polson2013}.
The expanded likelihood has a Gaussian form and can be combined with normal prior distributions on regression coefficients $\bm{\gamma}^{q}$ and an inverse-Wishart distribution on covariance matrix $\bm{\Sigma}^{q}$.
The parameters are known to have conditionally conjugate posterior distributions and allow for direct sampling from their multivariate normal and inverse-Wishart distributions respectively, resulting in MCMC chains of the joint posterior distribution in Equation \ref{eq:post}.
We also include a few comments on prior specification for the proposed Gibbs sampling procedure in the Supplemental material.

As an alternative to the proposed Gibbs sampling procedure, sampling from the posterior distribution(s) of multinomial logistic regression coefficients can theoretically be done with other standard MCMC-methods for non-conjugate prior-likelihood combinations, such as Metropolis-Hastings \parencites(e.g.,)(Ch. 3 and 5){Chib1998,Forster2003,Rossi2005} or Hamiltonian Monte Carlo \cite[e.g.,][]{Betancourt2017,Thomas2021,Betancourt2013} sampling algorithms.

\section{Transformation of posterior regression coefficients to the probability scale}\label{s:treatmentcomparison}

The output of the BMMLR model from Section \ref{s:model} is an MCMC sample of posterior multinomial regression coefficients. 
These regression coefficients reflect the importance of a predictor on a specific joint response combination and represent - in exponentiated form - the odds compared to reference category $Q$.
While these regression coefficients can be insightful in a truly multinomial research problem, they have no straightforward interpretation in multivariate treatment comparison where marginal effects on individual dependent variables play a central role \parencite{FDA2017}.

Transformation of regression coefficients to the multivariate probability scale forms a convenient solution to gain more intuitive insights in both joint and marginal treatment effects.
These transformations rely on the close relationship between multinomial and multivariate parametrizations and can be flexibly obtained for the trial population (i.e., average treatment effects) or for subpopulations (i.e., conditional average treatment effects).
They are directly suitable for statistical decision-making regarding treatment comparison.

We use the framework for transformation to the probability scale and decision-making with a posterior sample of multivariate treatment differences introduced in  \cite{Kavelaars2020} and \cite{Kavelaars2022}.
Technical details of these procedures are presented in Algorithm \ref{alg:procedure_sample} in Appendix \ref{app:transformation}. 
We use the remainder of this section to summarize and illustrate the procedure with a toy example from the IST-3-data, where we assume interest in the effect of Alteplase in the experimental condition ($T_{A}$) compared to no treatment in the control group ($T_{C}$).

Assume that we re-analyze a part of the IST-3 data using the BMMLR framework and take one of originally presented analyses as a starting point \parencite{IST2012}. 
In the selected analysis, the researchers compared the effects of Alteplase vs. control on their primary outcome, long-term independent living after six months ($Indep6$), among subgroups of patients based on the severity of their initial stroke.
In our example, we perform a multivariate analysis of the treatment effects on the primary outcome ($Indep6$) and one of the secondary (short-term) dependent variables: being stroke-free in the first seven days after the initial stroke ($Strk7$).
We incorporate severity of the initial stroke as a predictor variable to study heterogeneity, using the grouping criteria from the original trial for the estimation of conditional average treatment effects.
We aim to investigate the average treatment effect among the trial population as specified by the original eligibility criteria for inclusion.
We are also interested in a potential interaction between the treatment and stroke severity, and investigate the conditional average treatment effects among patients with various severities of stroke.
To take the clustered structure of the data into account, we specified a BMMLR mixed-effects model with random slopes for the intercept and the main treatment effect, resulting in the following linear predictor:
\begin{flalign}\label{eq:psi_app}
	\psi^{q}_{ji} &= \gamma^{q}_{0j} + \gamma^{q}_{1j} T_{ji} + \beta^{q}_{2} NIHSS_{ji} +  \beta^{q}_{3} NIHSS_{ji} T_{ji}\\\nonumber
	\gamma^{q}_{0j} & = \gamma^{q}_{00} + u_{0j}\\\nonumber
	\gamma^{q}_{1j} & = \gamma^{q}_{10} + u_{1j}.\nonumber
\end{flalign} 
\noindent In Equation \ref{eq:psi_app}, $\bm{x}_{ji} = (1,T_{ji}, NIHSS_{ji}, NIHSS_{ji}T_{ji})$ with treatment indicator $T_{ji}$ and $NIHSS_{ji}$ being the stroke severity score of subject $i$ in hospital $j$.
The $Q=4$ resulting joint response categories are $(\{Strk7 = 1, Indep6 = 1\}, \{Strk7 = 1, Indep6 = 0\}, \{Strk7 = 0, Indep6 = 1\}, \{Strk7 = 0, Indep6 = 0\})$, which we refer to as $(\{11\}, \{10\}, \{01\}, \{00\})$.

\subsection{Transformation to cluster-specific (differences between) probabilities}\label{sss:transformation}
The main quantity of interest, the (cluster-specific) marginal multivariate treatment difference, is defined as the difference between cluster-specific multivariate success probabilities of the two treatments:
\begin{flalign}\label{eq:delta_app}
	\delta^{Strk7}_{j} & = \theta_{Aj}^{Strk7} - \theta_{Cj}^{Strk7}\\\nonumber
	\delta^{Indep6}_{j} & = \theta_{Aj}^{Indep6} - \theta_{Cj}^{Indep6}\nonumber
\end{flalign} 
\noindent where subscripts $Aj$ and $Cj$ indicate cluster-specific parameters of the (experimental) Alteplase and control treatments respectively. The elements on the right-hand sides of Equation \ref{eq:delta_app}, success probabilities $\theta_{Tj}^{k}$, are sums of the multinomial joint response probabilities of all response categories with a success on outcome $k$:
\begin{flalign}\label{eq:theta_app}
	\theta_{Tj}^{Strk7} & = p(\bm{y}_{j} = \{11\}|T) + p(\bm{y}_{j} = \{10\}|T) = \phi_{Tj}^{1} + \phi_{Tj}^{2}\\
	\theta_{Tj}^{Indep6} & = p(\bm{y}_{j} = \{11\}|T) + p(\bm{y}_{j} = \{01\}|T) = \phi_{Tj}^{1} + \phi_{Tj}^{3}\nonumber
\end{flalign}
\noindent

The multinomial joint response probabilities $\bm{\phi}_{Tj}$ that form the elements of success probabilities $\bm{\theta}_{Tj}$ follow from plugging in posterior regression coefficients $\bm{\gamma}^q_{j}$ in the linear predictor (Equation \ref{eq:psi_app}) and the multinomial logistic link function (Equation \ref{eq:H_lik_mult}) for prespecified covariates $\bm{x}_{j}$ and for the relevant response category $q$.
\begin{flalign}\label{eq:phi_app}
	\phi^{q}_{Tj} = 
	& = \frac{\exp{(\psi^{q}_{Tj})
	}}{ \displaystyle\sum_{r=1}^{Q-1} \exp{(\psi^{r}_{Tj}
			)} + 1}. 
\end{flalign}
The information in covariate vector $\bm{x}_{j}$, which directly affects $\psi_{Tj}^q$, determines the treatment as well as the subpopulation of interest.  
Subpopulations can be defined as a value, such as a stroke severity score of one standard deviation below or above the mean, that can be plugged in directly into Equations \ref{eq:psi_app} and \ref{eq:H_lik_mult}. 
When interested in a subpopulation that is defined by an interval, such as the groups of stroke severity in the IST-3, the joint response probability is marginalized over the specified interval or averaged over a sample of observations in this interval. 
In the latter case, joint response probability $\phi^{q}_{Tj}$ is computed for each observed subject $i \in 1,\dots,n_{j}$ via Equation \ref{eq:H_lik_mult}.
The joint response probability for each treatment $T$ is then computed by averaging over all subjects $i$ in treatment $T$ and cluster $j$.

Since the model in Section \ref{s:model} resulted in a sample of $L$ posterior draws of each regression coefficient, multivariate treatment differences are computed for each draw $(l)$ separately. 
The resulting posterior samples can be summarized with standard descriptive methods.

\subsection{Pooling treatment effects over clusters}
As a last step, cluster-specific estimates are pooled into estimates of average or conditional treatment effects among (sub)populations of interest via the following procedure: 
\begin{flalign}\label{eq:pool}
	\bm{\delta} & = \frac{\displaystyle\sum_{j=1}^{J} n_{j} \bm{\delta}_{j}}{\displaystyle\sum_{j=1}^{J} n_{j}}  
\end{flalign}
This pooling strategy weighs cluster-specific estimates by cluster size, thereby balancing data with unequal cluster sizes.

\section{Decision-making based on multivariate treatment effects}\label{s:decision}

\begin{figure}
	\centering
	\includegraphics[width=0.6\linewidth,keepaspectratio]{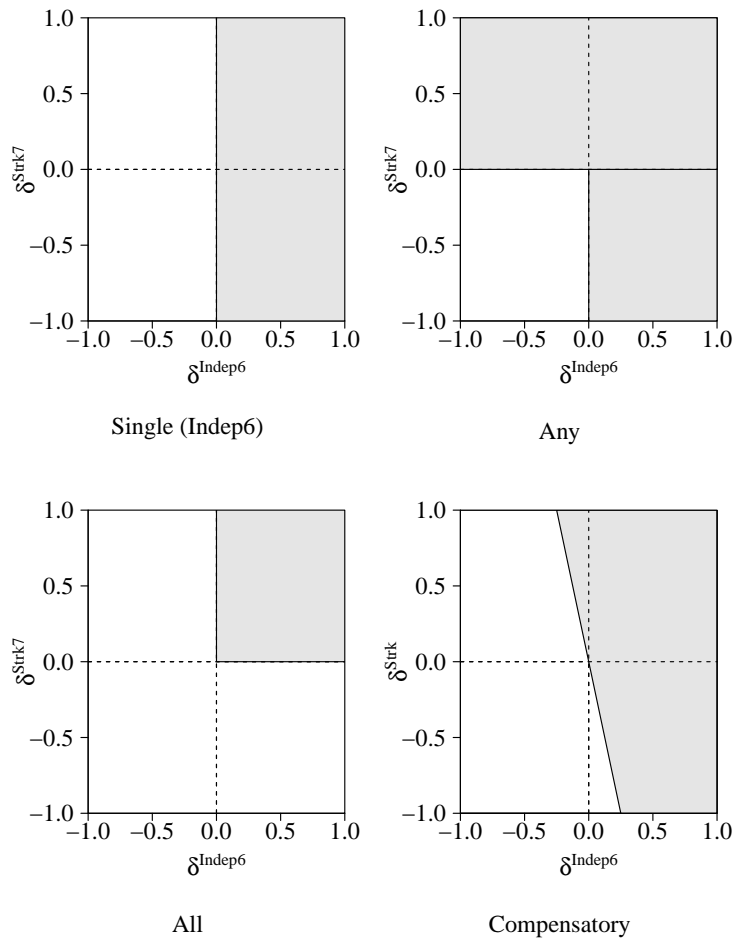}
	\caption{Superiority regions of four decision rules applied to the IST-3. 
		The Compensatory rule has weights $\bm{w}=(0.20,0.80)$.}
	\label{fig:decision}
\end{figure}

The obtained sample of posterior treatment differences can be used for statistical decision-making regarding treatment superiority and inferiority.
The multivariate context has multiple options to define superiority and inferiority, leaving much flexibility to combine and prioritize dependent variables in a suitable way.
We shortly discuss four different decision rules to give some idea of possibilities, without intending to be exhaustive or complete.
The presented rules have different theoretical underpinnings and distinct statistical properties, such as acceptance regions, a priori estimated sample sizes, cutoff values, and error rates.
The acceptance regions for superiority decisions of the four presented rules are graphically presented in Figure \ref{fig:decision}.
More details to guide an informed choice for one of these decision rules in practice can be found in Kavelaars et al. \cite{Kavelaars2020}.

Three of these rules originate from guidelines of the Food and Drug Administration (FDA) \cite{FDA2017}. 
The FDA defines superiority as a treatment difference larger than zero on the primary outcome (which we refer to as \enquote{Single rule}), on all dependent variables (\enquote{All rule}) or on any of the dependent variables (\enquote{Any rule}).
The Single rule reduces the statistical analysis to a univariate problem, using only the treatment difference of independent living after $6$ months as a primary outcome (Single rule). 
The All and Any rules make no distinction in the importance of dependent variables and assume that the short-term and long-term outcome are either both required for superiority or inferiority (All rule), or are interchangeable (Any rule).

In practice, these rules can oversimplify decision-making.
Secondary outcome variables often contribute to treatment evaluation as well, but are given a co-primary status in the All and Any rules or are not formally included in the statistical decision procedure when the Single rule is used \parencite[][]{Sozu2010,Sozu2016}. 
To handle outcomes that differ in relative importance, linear combinations of dependent variables with pre-assigned (importance) weights have been proposed as a flexible alternative \parencite{OBrien1984,Murray2016,Whitehead2010,Su2012,Kavelaars2020}. 
We refer to a linear combination as a Compensatory rule, referring to its inherent mechanism that allows (weighted) positive and negative effects to compensate each other.
The Compensatory rule allows the IST-3 data to consider the effects on the long-term much more important than the short-term effect without completely excluding the risk of a recurrent stroke from the final decision. 
In such a situation, we can assign the primary outcome ($Indep6$) - for example - four times more weight than the secondary outcome ($Strk7$) and consider Alteplase superior to no treatment if a lower chance of dependency is outweighed by a small increase in the risk of a recurrent stroke.

Evidence in favor of the decision rule can be quantified by the proportion posterior draws of the pooled treatment difference $\bm{\delta}$ that lie in the decision-rule specific acceptance region, denoted by $\mathcal{S}_{R}$.
A conclusion is reached via comparison to $p_{cut}$, which is a cutoff value to balance the required amount of evidence with anticipated Type I error rates \parencite{Marsman2016}:
\begin{flalign}\label{eq:superiority}
	p(\bm{\delta} \in \mathcal{S}_{R}) > p_{cut}.
\end{flalign}
In the multivariate logistic regression model, the probability in Equation \ref{eq:superiority} has no analytical solution. 
Therefore, decisions are made via the posterior MCMC-sample of $L$ draws.
Superiority is concluded when:
\begin{flalign}\label{eq:superiority_app}
	\frac{1}{L} \displaystyle\sum_{(l)=1}^{L} I(\bm{\delta}^{(l)} \in \mathcal{S}_{R}) > p_{cut}.
\end{flalign}
Similarly, inferiority is concluded when:
\begin{flalign}\label{eq:inferiority_app}
	\frac{1}{L} \displaystyle\sum_{(l)=1}^{L} I(\bm{\delta}^{(l)} \in \mathcal{S}_{R}) < 1 - p_{cut}.
\end{flalign}
In Section \ref{s:application}, we demonstrate these decision with data from the IST-3 as part of an illustration of the BMMLR framework.

\section{Numerical evaluation}\label{s:evaluation}
The current section presents an evaluation of the performance of the proposed BMMLR framework. 
The goal of the evaluation was twofold and we aimed to demonstrate:
\begin{enumerate}
	\item how well the obtained regression coefficients and treatment effects correspond to their true values to examine bias;
	\item how often the BMMLR framework results in an (in)correct superiority or inferiority conclusion to learn about decision error rates;
\end{enumerate}
\subsection{Setup}
\paragraph{Fitted models.}\label{sss:setup_fittedmodels}
The performance of the multilevel model was evaluated in a treatment comparison based on a two-level model with two dependent variables and one covariate at the subject level. 
We compared the method to two different (single-level) reference approaches, resulting in the following three modelling procedures:

\begin{enumerate}
	\item The BMMLR model presented in Section \ref{s:model}.
	We generated response data from a mixed effects model to include random effects while keeping the number of estimated parameters limited.
	We included an interaction between the treatment and the covariate as well, resulting in the following linear predictor:
	\begin{flalign}\label{eq:psi_eval}
		\psi^{q}_{ji} &= \gamma^{q}_{0j} + \gamma^{q}_{1j} T_{ji} + \beta^{q}_{2} w_{ji} +  \beta^{q}_{3} w_{ji} T_{ji}\\\nonumber
		\gamma^{q}_{0j} & = \gamma^{q}_{00} + u_{0j}\\\nonumber
		\gamma^{q}_{1j} & = \gamma^{q}_{10} + u_{1j}.\nonumber
	\end{flalign} 
	\noindent 
	In line with previous notation, $\bm{x}_{ji} = (1,T_{ji}, w_{ji}, w_{ji}T_{ji})$ in Equation \ref{eq:psi_eval}. 
	Further, vector $\bm{\gamma}^{q}_{j} = (\gamma^{q}_{0j}, \gamma^{q}_{1j})$ reflects random effects with multivariate normally distributed errors (i.e., $(u^{q}_{0j},u^{q}_{1j}) \sim N(\bm{0},\bm{\Sigma}^{q})$) for the intercept and main effect of the treatment.
	Regression coefficients $\bm{\beta}^{q} = (\beta^{q}_{2}, \beta^{q}_{3})$ reflect fixed effects for the covariate and covariate-by-treatment interaction.

	\item Single-level Bayesian multivariate logistic regression model \parencite[BMLR;][]{Kavelaars2022}, as a first reference approach.
	For this model, we use a restricted version of Equation \ref{eq:psi_eval} with fixed regression coefficients only:
	\begin{flalign}
		\psi^{q}_{ji} & = \beta^{q}_{0} + \beta^{q}_{1} T_{ji} + \beta^{q}_{2} w_{ji} + \beta^{q}_{3} w_{ji}T_{ji},
	\end{flalign} 
	\noindent 
	MCMC chains were sampled with a simplified version of the Gibbs sampling procedure in Appendix \ref{app:posterior_computation}, that iterates over $\bm{\beta}$ and $\bm{\Omega}$.
	The model shares information in the estimation of conditional treatment effects with sufficient power, but does not take the multilevel structure of the data into account.
	
	\item Single-level unconditional Bayesian multivariate Bernoulli analysis \parencite[BMB;][]{Kavelaars2020}, as a second reference approach.
	Bayesian multivariate Bernoulli analysis relies on a conjugate multinomial likelihood and Dirichlet prior.
	MCMC draws are sampled directly from the posterior Dirichlet distribution with parameters $\sum_{j=1}^{J} \sum_{i=1}^{n_{j}}  I(\bm{y}_{ji} = \bm{h}^{q}) + \alpha^{0q}$, where we assigned prior hyperparameters $\bm{\alpha}^{0} = (0.01, 0.01, 0.01, 0.01)$.
	The approach can estimate homogeneous treatment effects accurately and fast, but cannot deal with multilevel data.
	Moreover, conditional treatment effects originate from subsampling, which is less powerful than regression due to the isolation from other information.
\end{enumerate}

\paragraph{Effect size.}
We specified a heterogeneous treatment effect, with pooled average treatment differences of zero ($\bm{\delta} = (0,0)$, $\delta (\bm{w}) = 0$) and pooled conditional treatment differences larger than zero ($\bm{\delta} = (0.25,0.15)$, $\delta (\bm{w}) = 0.20$). 
This scenario aimed to demonstrate the Type I error rate among the trial population. 
It reflects a least favorable treatment difference for the Any and Compensatory rules and should therefore result in the targeted Type I error rate for these rules to be considered accurate.
The conditional treatment effect provided insight in the power to conclude superiority among the subpopulation under consideration.
Outcome variables were negatively correlated ($\rho_{ATE} = -.157$; $\rho_{CATE} = -.20$).	
The regression parameters used to generate these effects are presented in Table \ref{tab:EffectSize}.

\begin{table}[ht]
	\centering
	\captionsetup{justification=centering}
	\caption{True regression parameters used for data generation} 
	\label{tab:EffectSize}
	\begin{tabular}{lrrrr}
		\toprule
		& $q_{1}$ & $q_{2}$ & $q_{3}$ & $q_{4}$ \\ 
		\midrule
		$p_{0}$ (Intercept) &  0.000 &  0.433 &  0.433 &  0.000 \\ 
		$p_{1}$ ($T_{ji}$) &  0.000 &  0.000 &  0.000 &  0.000 \\ 
		$p_{2}$ ($w_{ji}$) &  1.027 &  0.601 &  0.427 &  0.000 \\ 
		$p_{3}$ ($w_{ji}T_{ji}$)& -2.055 & -1.201 & -0.854 &  0.000 \\ 
		\bottomrule
	\end{tabular}
\end{table}

For the BMMLR model, the covariance matrix of random effects, $\bm{\Sigma}^{q}$, was specified as:
\begin{flalign}
	\begin{bsmallmatrix}
		0.1 & 0 \\
		0 & 0.1 \\
	\end{bsmallmatrix}
\end{flalign}
\noindent for all $q \in 1,\dots,Q-1$.

\paragraph{Sample size.}
We varied the sample sizes at the cluster and subject level.
Since there are no clear guidelines regarding sample size computations in multilevel multivariate logistic regression, we explored performance of the model for different numbers of clusters and different sample sizes within clusters. 
Specifically, we used number of clusters $J \in \{10,100\}$ and observations per cluster $n_{j} \in \{10,100\}$ for each treatment, resulting in four different sample size combinations.

\subsubsection{Procedure}

\paragraph{Data generation.}
For each sample size, we sampled $1000$ datasets under the mixed effects model in Equation \ref{eq:psi_eval} with the true regression parameters in Table \ref{tab:EffectSize}.
We assigned $n_{j}$ participants to each treatment $T$ and generated covariate $x$ from a standard normal distribution.
We sampled response vector $\bm{y}_{ji}$ from a multinomial distribution with probabilities $\bm{\phi}_{ji}$.

\paragraph{Gibbs sampling.}
Regression coefficients for the BMMLR and BMLR models were estimated via the Gibbs sampling procedure in Appendix \ref{app:posterior_computation}. 
We ran two MCMC-chains via the Gibbs sampler introduced in Section \ref{s:model} with $L = 50,000$ iterations plus $10,000$ burn-in iterations. 
This large number of iterations aims to minimize the influence of the potentially high autocorrelations between parameters in multilevel models on the stationary distribution of the parameters.
Autocorrelations were highest among random effect parameters $\bm{\gamma}_{j}$ and ranged between $0.107$ and $0.781$ at lag $1$ and reduced to a range of $-0.012 - 0.276$ at lag $10$. 
Further, following the guidelines in Gelman et al. \cite{Gelman2013}, we ensured that the multivariate potential scale reduction factor was below $1.10$. 

\paragraph{Prior specification.}\label{par:evaluation_prior}
For the multilevel model (BMMLR), we specified diffuse priors, which were multivariate normally distributed for regression coefficients:
\begin{flalign}\label{eq:prior}
(\beta^{q}_{2}, \beta^{q}_{3}) & \sim N(\begin{bsmallmatrix}
	0 \\
	0 \\
\end{bsmallmatrix}, 
\begin{bsmallmatrix}
	10 & 0 \\
	0 & 10 \\
\end{bsmallmatrix})\\\nonumber
(\gamma^{q}_{00}, \gamma^{q}_{10}) & \sim N(\begin{bsmallmatrix}
	0 \\
	0 \\
\end{bsmallmatrix}, 
\begin{bsmallmatrix}
	10 & 0 \\
	0 & 10 \\
\end{bsmallmatrix})\nonumber
\end{flalign}
\noindent The specified variance matrices of regression coefficients were motivated by a paper of Gelman et al. \cite{Gelman2008}, who recommend to choose a variance parameter that results in realistic support for the probability parameter after non-linear transformation in logistic regression. 
We specified an inverse-Wishart prior distribution for the covariance matrix:
\begin{flalign*}
\bm{\Sigma}^{q} & \sim \mathcal{W}^{-1} (2, \begin{bsmallmatrix}
	0.1 & 0 \\
	0 & 0.1 \\
\end{bsmallmatrix}).\nonumber
\end{flalign*} 
\noindent The regression parameters $\bm{\beta}^{q}$ in the single-level regression model (BMLR) were the same as in the multivariate approach (i.e., independent normal priors with means of 0 and variances of 10).

\paragraph{Transformation and decision-making.}\label{sss:decision_rules}
We applied the procedures in Algorithm \ref{alg:procedure_sample} to use the obtained MCMC-chains of posterior regression coefficients for superiority decision-making. 
We thinned the chains in the transformation procedure with a factor $10$ to reduce the computational burden.

We considered two different effects: 
\begin{enumerate}
\item an average treatment effect for the trial population;
\item a conditional treatment effect for a subpopulation scoring one standard deviation below the mean or lower;
\end{enumerate}  
The treatment effects required marginalization over the interval that defined the (sub)population, which we accomplished by averaging over joint response probabilities computed for the empirical sample of data. 
Cluster-specific treatment effects were weighed by their sample sizes to produce a pooled estimate of the treatment difference.

Decisions were made with a right-sided test for the All, Any, and Compensatory (equal weights, $\bm{w} = (0.50,0.50)$) rules with formal superiority regions:
\begin{enumerate}
\item Any rule:
$\mathcal{S}_{R}
= \{ \bm{\delta} | \max_{1<k<K} \delta^{k} > 0\}  | \bm{y}, \bm{x}$ and cut-off value $p_{cut} = 1 - \frac{\alpha}{K}$
\item All Rule:
$\mathcal{S}_{R} 
= \{ \bm{\delta} | \min_{1<k<K} \delta^{k} > 0\}  | \bm{y}, \bm{x}$ and cut-off value $p_{cut} = 1 - \alpha$
\item Compensatory rule:
$\mathcal{S}_{R} 
= \{ \bm{\delta} | \delta(\bm{w}) > 0\} | \bm{y}, \bm{x}$ and cut-off value $p_{cut} = 1 - \alpha$
\end{enumerate}

We computed the probability to conclude superiority ($p_{Sup}$) as the proportion of posterior treatment differences in the superiority region via Equation \ref{eq:superiority}.
The targeted Type I-error rate of $\alpha = .05$ corresponded to decision threshold $p_{cut} = 1 - \alpha = 0.95$ (Compensatory and All rules) and a for multiple tests corrected threshold $p_{cut} = 1 - \frac{\alpha}{K} = 0.975$ (Any rule) \parencite{Marsman2016, Kavelaars2020, Sozu2016}.

\subsubsection{Software}
We conducted our analyses in \texttt{R} and made use of several existing packages \cite{RCT2020}.
P\'olya-Gamma variables were drawn with the \texttt{pgdraw} package \cite{Makalic2016}.
Further, we drew variables from the multivariate normal, truncated normal, and Dirichlet distributions with the \texttt{MASS}, \texttt{msm}, and \texttt{MCMCpack} packages respectively \cite{Venables2002,Jackson2011,Martin2011}.
MCMC chains were diagnosed with the \texttt{coda} and \texttt{mcmcse} packages \cite{Plummer2006,Flegal2021}. 
We parallellized the simulation procedure with the \texttt{foreach} and \texttt{doParallel} packages \cite{Microsoft2020,Microsoft2020a} and created \LaTeX{} tables with the \texttt{xtable} package \cite{Dahl2019}.
The \texttt{R} code used to generate results can be found on GitHub \url{https://github.com/XynthiaKavelaars/Bayesian-multilevel-multivariate-logistic-regression}.

\subsection{Results}

The current subsection presents the results of the simulation study. 
Presented decision error rates are in Table \ref{tab:pReject}.

\subsubsection{Bias}
Regression coefficients, variance matrices and treatment effects (success probabilities, treatment differences) could be estimated without bias in all sample sizes and data generating mechanisms.
The absolute average deviation of mean point estimates from true values was smaller than $.01$.

\subsubsection{Decision error rates}

\paragraph{Type I error rates}
The average treatment effect demonstrated that the probability to incorrectly conclude superiority in multilevel regression (BMMLR) was close to the targeted $.05$ under a least favorable scenario (i.e., Any and Compensatory decision rules). 
In general, both reference approaches suffered from inflated Type I error to a similar extent.

The amount of inflation in BMMLR was affected by sample size: A large number of clusters ($J=100$) and/or a large subjects per cluster ($n_{j}=100$) had the largest Type I error rates, with the combination $J=100,n_{j}=100$ resulting in the most severe inflation. 
On the other hand, a small number of clusters and a small number of subjects per cluster ($J=10, n_{j}=10$) resulted in an acceptable Type I error rate for the single-level BMLR model as well, suggesting some robustness against the violation of the assumption of independent observations in the current setup.
In general, the number of subjects per cluster appeared more influential on the Type I error rate inflation than the number of clusters, as demonstrated by the two scenarios with an identical total sample size ($J=10, n_{j}=100$ and $J=100, n_{j}=10$): A small number of clusters and a large sample size per cluster resulted in larger Type I error rates than a large number of clusters with a small sample size per cluster.
Keeping everything else constant, a larger number of clusters meant more independent units, implying that the assumption of independent observations was violated less severely.
In other words, the need for a multilevel model was more prominent when the number of clusters was small.
A similar pattern was seen under the All rule, although Type I errors were small in general. 
This was expected, since a) the All rule is known to be the most conservative of the three introduced rules; and b) the treatment difference was smaller than the least favorable scenario of this decision rule.

\paragraph{Power}
The conditional treatment effect demonstrated the power to correctly conclude superiority for all three rules.
Three results were highlighted.  
First, the multilevel model (BMMLR) is more powerful when the number of clusters is higher. 
The two conditions with an equal total sample size showed a $.30$ difference in power under the All rule.
The other rules showed the same patterns, but had too high proportions of superiority conclusions to clearly distinguish the sample size conditions: The power in the other conditions equaled or was close to the maximum of $1.000$.

Second, the single-level regression model (BMLR) resulted in more superiority conclusions than the multilevel regression model, implying that the posterior distributions of treatment differences of the single-level regression model had smaller variances. 
Again, differences were best illustrated by the All rule and the condition with small sample sizes for the Any and Compensatory decision rules, as these proportions were well below the maximum.  
Similar to the Type I error rates, the differences between the proportions of superiority conclusions appeared to be subject to the number of clusters, as demonstrated by a comparison of the two conditions with an identical total sample size under the All rule. 
The multilevel model was less powerful than the single-level model when the number of clusters was low in particular, being in line with non-independence of clustered observations.

Third, the multivariate Bernoulli model (BMB) has low power overall, despite the underestimation of variance due to falsely assuming independent observations. 
As a subsampling approach, conditional treatments were fitted on the part of the data that makes up the subpopulation of interest. 
Especially the $J=10$, $n_{j}=10$ condition suffered from a small remaining sample size.

\begin{table}[htbp]
\centering
\caption{Proportions of superiority decisions and standard errors by data-generating mechanism, estimation method, and decision rule.} 
\label{tab:pReject}
\begin{tabular}{lp{0.02cm}rrp{0.02cm}rrp{0.02cm}rr}
	\hline\noalign{\smallskip}
	\multicolumn{10}{c}{\textbf{Average treatment effect: }$\bm{\delta} = (0.000, 0.000)$, $\bm{\delta} (\bm{w}) = 0.000$} \\ 
	\noalign{\smallskip}\hline\noalign{\smallskip} 
	& & \multicolumn{2}{l}{Any}& & \multicolumn{2}{l}{All}& & \multicolumn{2}{l}{Compensatory}\\
	$J$ = 10, $n_{j}$ = 10 &  & \multicolumn{1}{l}{p} & \multicolumn{1}{l}{se} &  & \multicolumn{1}{l}{p} & \multicolumn{1}{l}{se} &  & \multicolumn{1}{l}{p} & \multicolumn{1}{l}{se} \\
	\noalign{\smallskip}\hline\noalign{\smallskip}
	BMMLR &   & $0.032$ & (0.006) &   & $0.000$ & (0.000) &   & $0.042$ & (0.006) \\ 
	BMLR &   & $0.055$ & (0.007) &   & $0.001$ & (0.001) &   & $0.059$ & (0.007) \\ 
	BMB &   & $0.050$ & (0.007) &   & $0.001$ & (0.001) &   & $0.046$ & (0.007) \\ 
	\noalign{\smallskip}\hline\noalign{\smallskip} 
	$J$ = 100, $n_{j}$ = 10& \multicolumn{9}{l}{ } \\ 
	\noalign{\smallskip}\hline\noalign{\smallskip} 
	BMMLR &   & $0.053$ & (0.007) &   & $0.002$ & (0.001) &   & $0.048$ & (0.007) \\ 
	BMLR &   & $0.077$ & (0.008) &   & $0.003$ & (0.002) &   & $0.066$ & (0.008) \\ 
	BMB &   & $0.069$ & (0.008) &   & $0.002$ & (0.001) &   & $0.056$ & (0.007) \\ 
	\noalign{\smallskip}\hline\noalign{\smallskip} 
	$J$ = 10, $n_{j}$ = 100& \multicolumn{9}{l}{ } \\ 
	\noalign{\smallskip}\hline\noalign{\smallskip} 
	BMMLR &   & $0.044$ & (0.006) &   & $0.000$ & (0.000) &   & $0.060$ & (0.008) \\ 
	BMLR &   & $0.200$ & (0.013) &   & $0.004$ & (0.002) &   & $0.125$ & (0.010) \\ 
	BMB &   & $0.188$ & (0.012) &   & $0.003$ & (0.002) &   & $0.113$ & (0.010) \\ 
	\noalign{\smallskip}\hline\noalign{\smallskip} 
	$J$ = 100, $n_{j}$ = 100& \multicolumn{9}{l}{ } \\ 
	\noalign{\smallskip}\hline\noalign{\smallskip} 
	BMMLR &   & $0.057$ & (0.007) &   & $0.000$ & (0.000) &   & $0.054$ & (0.007) \\ 
	BMLR &   & $0.252$ & (0.014) &   & $0.005$ & (0.002) &   & $0.169$ & (0.012) \\ 
	BMB &   & $0.245$ & (0.014) &   & $0.005$ & (0.002) &   & $0.159$ & (0.012) \\ 
	\noalign{\smallskip}\hline\noalign{\smallskip} 
	\multicolumn{10}{c}{\textbf{Conditional treatment effect: }$\bm{\delta} = (0.116, 0.069)$, $\bm{\delta} (\bm{w}) = 0.092$} \\ 
	\noalign{\smallskip}\hline\noalign{\smallskip} 
	& & \multicolumn{2}{l}{Any}& & \multicolumn{2}{l}{All}& & \multicolumn{2}{l}{Compensatory}\\
	$J$ = 10, $n_{j}$ = 10 &  & \multicolumn{1}{l}{p} & \multicolumn{1}{l}{se} &  & \multicolumn{1}{l}{p} & \multicolumn{1}{l}{se} &  & \multicolumn{1}{l}{p} & \multicolumn{1}{l}{se} \\
	\noalign{\smallskip}\hline\noalign{\smallskip} 
	BMMLR &   & $0.731$ & (0.014) &   & $0.245$ & (0.014) &   & $0.920$ & (0.009) \\ 
	BMLR &   & $0.397$ & (0.015) &   & $0.065$ & (0.008) &   & $0.587$ & (0.016) \\ 
	BMB &   & $0.183$ & (0.012) &   & $0.025$ & (0.005) &   & $0.294$ & (0.014) \\ 
	\noalign{\smallskip}\hline\noalign{\smallskip} 
	$J$ = 100, $n_{j}$ = 10& \multicolumn{9}{l}{ } \\ 
	\noalign{\smallskip}\hline\noalign{\smallskip} 
	BMMLR &   & $1.000$ & (0.000) &   & $0.995$ & (0.002) &   & $1.000$ & (0.000) \\ 
	BMLR &   & $1.000$ & (0.000) &   & $0.868$ & (0.011) &   & $1.000$ & (0.000) \\ 
	BMB &   & $0.933$ & (0.008) &   & $0.520$ & (0.016) &   & $0.980$ & (0.004) \\ 
	\noalign{\smallskip}\hline\noalign{\smallskip} 
	$J$ = 10, $n_{j}$ = 100& \multicolumn{9}{l}{ } \\ 
	\noalign{\smallskip}\hline\noalign{\smallskip} 
	BMMLR &   & $1.000$ & (0.000) &   & $0.949$ & (0.007) &   & $1.000$ & (0.000) \\ 
	BMLR &   & $0.997$ & (0.002) &   & $0.771$ & (0.013) &   & $1.000$ & (0.000) \\ 
	BMB &   & $0.917$ & (0.009) &   & $0.445$ & (0.016) &   & $0.969$ & (0.005) \\ 
	\noalign{\smallskip}\hline\noalign{\smallskip} 
	$J$ = 100, $n_{j}$ = 100& \multicolumn{9}{l}{ } \\ 
	\noalign{\smallskip}\hline\noalign{\smallskip} 
	BMMLR &   & $1.000$ & (0.000) &   & $1.000$ & (0.000) &   & $1.000$ & (0.000) \\ 
	BMLR &   & $1.000$ & (0.000) &   & $1.000$ & (0.000) &   & $1.000$ & (0.000) \\ 
	BMB &   & $1.000$ & (0.000) &   & $1.000$ & (0.000) &   & $1.000$ & (0.000) \\ 
	\bottomrule
	\multicolumn{10}{l}{p = proportion of superiority decisions}\\
	\multicolumn{10}{l}{se = Standard errors}\\
	\noalign{\smallskip}\hline
\end{tabular}
\end{table}

\section{Illustration with IST-3 data}\label{s:application}

To illustrate the proposed framework with real data, we re-analyzed a subset of data from the Third International Stroke Trial using the BMMLR framework  \parencite{IST2012,Sandercock2016}.
The included $3,035$ subjects in the IST-3 were recruited from $156$ different hospitals in $12$ different countries, resulting in multilevel data from patients clustered within hospitals and hospitals clustered within countries.
We selected a two-level subset of $1,447$ subjects from $75$ hospitals in the United Kingdom with a known health and survival status at six months after the initial stroke and a known or predicted severity score of the initial stroke (NIH Stroke Score; NIHSS) at randomisation. 
The cluster sizes were skewed and ranged from $1$ to $117$, with a median cluster size of $7$ (SD: $26.66$). 
Of the selected subset of data,  $n_{A} = 716$ subjects were in the Alteplase group (treatment = $1$) and $n_{C} = 731$ subjects were in the control group (treatment = $0$). 
We compared the effects of the two treatments on a) being stroke-free for seven days (0 = no; 1 = yes) and b) long-term independent living at six months (0 = no, 1 = yes), while taking the severity of the initial stroke into account. 
The NIHSS can range from $0$ to $42$ with a higher score indicating a more severe stroke. 
The average stroke severity score in the IST-3 was $13.12$ (SD: $6.91$) and comparable in both treatment groups.

\subsection{Method}
We fitted our model with random slopes for the intercept and the treatment effect.
We sought to compare our multilevel model (BMMLR) to the two single-level models (BMLR and BMB) from the \nameref{s:evaluation} section in treatment comparison of Alteplase and control on dependency after six months ($\delta^{Indep6}$) and recurrent stroke within seven days ($\delta^{Strk7}$).
The multilevel model (BMMLR) was fitted with the linear predictor in Equation \ref{eq:psi_app} and the linear predictor of the single-level regression model (BMLR) was:
\begin{flalign}\label{eq:psi_app_NH}
\psi^{q}_{ji} &= \beta^{q}_{0} + \beta^{q}_{1} T_{ji} + \beta^{q}_{2} NIHSS_{ji} +  \beta^{q}_{3} NIHSS_{ji} T_{ji}
\end{flalign} 

\paragraph{Prior specification.}\label{par:illustration_prior}
For the regression coefficients in the multilevel model (BMMLR) and the single-level regression model (BMLR), we specified independent normal prior distributions with means of $0$ and variances of $10$.
For covariance matrix $\Sigma^{q}$, we specified an improper uniform prior for the random effects covariance matrix for each category $q$, to enable testing for the presence of random effects in the model comparison step discussed later.

\paragraph{Gibbs sampling.}
We ran two MCMC-chains via the Gibbs samplers. 
Since the chains of regression coefficients were highly autocorrelated in the multilevel model (lag 10: $\bm{\beta}: 0.47 - 0.59$; $\bm{\gamma}: 0.62 - 0.80$, $\bm{\Sigma}: -0.01-0.38$), we sampled a large number of $500,000$ iterations plus $10,000$ burnin iterations. 
The multivariate potential scale reduction factor was below $1.01$ for all parameters, implying that there were no signals of non-convergence.
We thinned MCMC-chains in follow-up posterior transformations with a factor $10$ to reduce computational demands, resulting in inference based on $L = 50,000$ draws. 

\paragraph{Transformation and decision-making.}
We applied the procedures in Algorithm \ref{alg:procedure_sample} to the thinned MCMC-chains of posterior regression coefficients to make superiority decisions. 
We considered (conditional) average treatment effects among seven different (sub)populations: 
\begin{enumerate}
\item ATE: average treatment effects for all patients in the trial population;
\item CATE - Low range: conditional average treatment effects for patients with a stroke severity score between $0$ and $5$;
\item CATE - Mid-Low range: conditional average treatment effects for patients with a stroke severity score between $6$ and $14$;
\item CATE - Mid-High range: conditional average treatment effects for patients with a stroke severity score between $15$ and $24$;
\item CATE - High range: conditional average treatment effects for patients with a stroke severity score above $25$;
\item CATE - Low value: conditional treatment effects for patients with a stroke severity score of $5.18$, corresponding to $1$ standard deviation below the mean;
\item CATE - High value: conditional treatment effects for patients with a stroke severity score of $19.03$, corresponding to $1$ standard deviation above the mean.
\end{enumerate}  
The grouping criteria for CATEs of ranges were taken from the original IST-3 paper \parencite{IST2012}.

We performed two-sided tests for the All, Any, and Compensatory rules. 
Similar to the IST-3, we used living independently as the most important outcome in the Compensatory rule and specified weights $\bm{w} = (0.20,0.80)$ for remaining free of strokes and independent living respectively.
This specification implied that the long-term outcome had four times more impact on the decision than the short-term outcome. 
The targeted two-sided Type I-error rate of $\alpha = .05$ corresponded to decision threshold $p_{cut} = 1 - \frac{\alpha}{2} = 0.975$ (Compensatory and All rules) and a for multiple tests corrected threshold $p_{cut} = 1 - \frac{\alpha}{2K} = 0.9875 $ (Any rule).

\subsubsection{Model comparison}\label{sss:method_modelcomparison}
Since the true model of these real-world data is unknown, we followed up on the analysis with a comparison of model fit via Bayes factors.
Bayes factors \cite{Jeffreys1961} quantify the relative evidence in the data between competing statistical models. Here we use default Bayes factors which avoid the need to manually specify prior distributions \cite{Mulder2021,Mulder2019,Vieira2023}.

\paragraph{BMLR vs. BMB.}
To compare the two single-level models, we computed a Bayes factor on the probabilities that the regression coefficients of the covariate ($\beta^{q}_{2}$) and the interaction between the covariate and the interaction ($\beta^{q}_{3}$) was equal to zero for all $q \in q,\dots,Q-1$ using the \texttt{BF()}-function from the \texttt{R}-package \texttt{BFpack} \parencite{Mulder2021}.

\paragraph{BMMLR vs. BMLR.}
To compare the proposed multilevel model (BMMLR) and the single-level model (BMLR), we computed empirical Bayes factors as proposed by Vieira-Generoso et al. \cite{Vieira2023}, which tests whether the random effects are equal across clusters using uniform priors for the random effects covariance matrices. 
This test is executed separately for all six different random effects in the multilevel model.

\paragraph{Software.}
In addition to the software packages used in Section \ref{s:evaluation}, we used \texttt{R} packages \texttt{haven} to import the dataset \parencite{Wickham2021}, \texttt{BFpack} \parencite{Mulder2021} to compute Bayes factors for comparison of the two single-level models.

\subsection{Results}

\begin{table}[htbp]
\centering
\caption{Average (ATE) and conditional average (CATE) treatment effects of the specified (sub)populations of the IST-3.} 
\label{tab:App}
\begin{tabular}{lp{0.02cm}rrrrp{0.02cm}rrr}
	\hline\noalign{\smallskip}
	& & ($\delta^{Strk7}, \delta^{Indep6}$) & Pop & Any & All & & $\delta (\bm{w})$ & Pop & Comp \\
	\noalign{\smallskip}\hline\noalign{\smallskip} 
	\multicolumn{3}{l}{ATE} & \multicolumn{7}{l}{$n_{A} = 716$, $n_{C} = 731$ } \\ 
	\noalign{\smallskip}\hline\noalign{\smallskip}
	BMMLR &   & ($-0.114,  0.029$) & ($0.000, 0.886$) & $\bm{<}$ & - &   & $ 0.000$ & $0.504$ & - \\ 
	BMLR &   & ($-0.116,  0.033$) & ($0.000, 0.941$) & $\bm{<}$ & - &   & $ 0.003$ & $0.572$ & - \\ 
	BMB &   & ($-0.117,  0.032$) & ($0.000, 0.911$) & $\bm{<}$ & - &   & $ 0.003$ & $0.549$ & - \\ 
	\noalign{\smallskip}\hline\noalign{\smallskip} 
	\multicolumn{3}{l}{CATE - Low range } & \multicolumn{7}{l}{$n_{A} = 99$, $n_{C} = 105$ } \\ 
	\noalign{\smallskip}\hline\noalign{\smallskip} 
	BMMLR &   & ($-0.078, -0.023$) & ($0.003, 0.317$) & $\bm{<}$ & - &   & $-0.034$ & $0.200$ & - \\ 
	BMLR &   & ($-0.081, -0.016$) & ($0.004, 0.365$) & $\bm{<}$ & - &   & $-0.029$ & $0.225$ & - \\ 
	BMB &   & ($-0.110, -0.036$) & ($0.019, 0.318$) & - & - &   & $-0.051$ & $0.207$ & - \\ 
	\noalign{\smallskip}\hline\noalign{\smallskip} 
	\multicolumn{3}{l}{CATE - Mid-Low range } & \multicolumn{7}{l}{$n_{A} = 327$, $n_{C} = 334$ } \\ 
	\noalign{\smallskip}\hline\noalign{\smallskip} 
	BMMLR &   & ($-0.090,  0.038$) & ($0.000, 0.884$) & $\bm{<}$ & - &   & $ 0.013$ & $0.679$ & - \\ 
	BMLR &   & ($-0.092,  0.044$) & ($0.000, 0.937$) & $\bm{<}$ & - &   & $ 0.017$ & $0.752$ & - \\ 
	BMB &   & ($-0.114,  0.045$) & ($0.001, 0.853$) & $\bm{<}$ & - &   & $ 0.013$ & $0.642$ & - \\ 
	\noalign{\smallskip}\hline\noalign{\smallskip} 
	\multicolumn{3}{l}{CATE - Mid-High range } & \multicolumn{7}{l}{$n_{A} = 237$, $n_{C} = 252$ } \\ 
	\noalign{\smallskip}\hline\noalign{\smallskip} 
	BMMLR &   & ($-0.139,  0.051$) & ($0.000, 0.992$) & $\bm{<} \& \bm{>}$ & - &   & $ 0.013$ & $0.753$ & - \\ 
	BMLR &   & ($-0.141,  0.054$) & ($0.000, 0.995$) & $\bm{<} \& \bm{>}$ & - &   & $ 0.015$ & $0.783$ & - \\ 
	BMB &   & ($-0.118,  0.047$) & ($0.006, 0.938$) & $\bm{<}$ & - &   & $ 0.014$ & $0.694$ & - \\ 
	\noalign{\smallskip}\hline\noalign{\smallskip} 
	\multicolumn{3}{l}{CATE - High range } & \multicolumn{7}{l}{$n_{A} = 53$, $n_{C} = 40$ } \\ 
	\noalign{\smallskip}\hline\noalign{\smallskip} 
	BMMLR &   & ($-0.183,  0.020$) & ($0.002, 0.980$) & $\bm{<}$ & - &   & $-0.021$ & $0.100$ & - \\ 
	BMLR &   & ($-0.188,  0.021$) & ($0.001, 0.982$) & $\bm{<}$ & - &   & $-0.021$ & $0.100$ & - \\ 
	BMB &   & ($-0.173,  0.019$) & ($0.069, 0.687$) & - & - &   & $-0.019$ & $0.327$ & - \\ 
	\noalign{\smallskip}\hline\noalign{\smallskip} 
	\multicolumn{3}{l}{CATE - Low value } & \multicolumn{7}{l}{ } \\ 
	\noalign{\smallskip}\hline\noalign{\smallskip} 
	BMMLR &   & ($-0.078, -0.007$) & ($0.002, 0.440$) & $\bm{<}$ & - &   & $-0.021$ & $0.291$ & - \\ 
	BMLR &   & ($-0.080,  0.000$) & ($0.002, 0.503$) & $\bm{<}$ & - &   & $-0.016$ & $0.328$ & - \\ 
	\noalign{\smallskip}\hline\noalign{\smallskip} 
	\multicolumn{3}{l}{CATE - High value } & \multicolumn{7}{l}{ } \\ 
	\noalign{\smallskip}\hline\noalign{\smallskip} 
	BMMLR &   & ($-0.140,  0.052$) & ($0.000, 0.991$) & $\bm{<} \& \bm{>}$ & - &   & $ 0.014$ & $0.751$ & - \\ 
	BMLR &   & ($-0.142,  0.055$) & ($0.000, 0.994$) & $\bm{<} \& \bm{>}$ & - &   & $ 0.015$ & $0.777$ & - \\ 
	\midrule
	\multicolumn{10}{l}{Pop = Posterior probability}\\ 
	\multicolumn{10}{l}{$>$ = superiority concluded}\\
	\multicolumn{10}{l}{$<$ = inferiority concluded}\\
	\noalign{\smallskip}\hline
\end{tabular}
\end{table}

\subsubsection{Results of different (sub)populations}
Table \ref{tab:App}  
show how different analysis models and different decision rules provide elaborate insights in the effects of Alteplase vs. control on a combination of dependent variables among different (sub)populations.
Analysis of the selected data with the BMMLR, BMLR, and BMB models gave the following results.

\paragraph{Average treatments effects.}
The average treatment effect (ATE) among the UK-based part of the trial population showed that the Alteplase group had a lower estimated probability of remaining free of strokes, a higher estimated probability of living independently, and a weighted probability difference close to zero.
The three modelling procedures produced similar estimates and unanimously resulted in the conclusions that Alteplase was inferior according to the Any rule due to the effect on being free of strokes, while neither superiority nor inferiority could be concluded from the All or Compensatory rules. 

\paragraph{Conditional average treatment effects.}
The four conditional average treatment effects (CATEs) that reflected subpopulations as ranges sketched a more heterogeneous picture than the average treatment effects. 
Whereas all ranges showed a lower probability of being free of strokes after treatment with Alteplase, these probabilities increased with the severity of the stroke. 
Differences between success probabilities of the two treatments appeared to increase with severity of the stroke, such that Alteplase appeared to have the largest negative effect on being stroke-free when the severity of the initial stroke was highest.
A more diffuse relation between stroke severity and treatment difference emerged on long-term independent living. 
Alteplase resulted in a slightly lower point estimate of the probability of independent living among patients with a Low stroke severity, but resulted in a higher estimated probability of independent living in all categories of more severe strokes.
Patients in the Mid-Low and Mid-High ranges of stroke severity had the largest positive effect of Alteplase on independent living.
The Low and High stroke severity patients had slightly higher weighted probabilities after Alteplase, while patients with a Mid-Low and Mid-High stroke severity had weighted probabilities close to zero.
These non-zero point estimates were not unanimously supported by sufficient evidence to conclude superiority or inferiority.
The All and Compensatory rules remained inconclusive for all models among all subpopulations.
The BMMLR and BMLR were unanimous in their conclusions for the Any rule:
Inferiority was concluded for patients with a Low, Mid-Low and High stroke severity, while both superiority and inferiority were concluded for patients with a Mid-High range stroke severity.
The BMB model remained inconclusive in the Low and High ranges and concluded inferiority among patients with a Mid-Low or Mid-High stroke severity, according to the Any rule.

The two conditional average treatment effects (CATEs) that specified subpopulations by values illustrated treatment differences for two hypothetical individual patients. 
After receiving Alteplase, both patients would have a lower probability of remaining free of strokes. 
Only the patient with a High stroke severity value had a higher probability of long-term independent living. 
The weighted failure probability difference was slightly below zero for the patient with a Low stroke severity and around zero for the patient with a High stroke severity. 
Again, the All and Compensatory rules remained inconclusive, whereas the Any rule would result in an inferiority conclusion for the patient with a Low stroke severity and in both inferiority and superiority for the patient with a High stroke severity.

\paragraph{Model comparison}\label{sss:results_modelcomparison}

Bayes factors \parencite{Kass1995} are computed to test whether there is evidence that a dependency structure is present in the data that is caused by the multilevel structure. 
The results are presented in Table \ref{tab:BF}.
These results indicate that there is evidence that each of the six different random effects do not vary across clusters. 
This implies that the parsimonious single-level model (BMLR) is preferred over the multilevel model (BMMLR) for these specific data. 
This result is also in agreement with the obtained estimates which are virtually identical under both models.
Model comparison between the two single-level models (BMLR vs. BMB) resulted in a log-transformed Bayes factor of $16.348$, reflecting strong evidence that the a regression model (BMLR) fitted the data better than the multivariate Bernoulli (BMB) model. 
We give a general recommendation on model selection in the Discussion section.

\begin{table}[ht]
\centering
\caption{Logarithmic transformations of Bayes factors of BMLR vs. BMMLR} 
\label{tab:BF}
\begin{tabular}{lrrr}
	\toprule
	& $q = 1$ & $q = 2$ & $q = 3$ \\ 
	\midrule
	NIHSS & 5.769 & 5.642 & 11.238 \\ 
	NIHSS $\times$ Trt & 5.653 & 6.181 & 8.555 \\ 
\end{tabular}
\end{table}

\subsubsection{Conclusions and discussion}
Several conclusions regarding the BMMLR framework could be drawn from the presented results.
First, multilevel analysis did not affect point estimates in the used subset of IST-3 data: BMMLR and BMLR models resulted in similar point estimates of $\bm{\delta}$ and $\delta (\bm{w})$, as expected from the negligible bias in the results of the simulation study.
The posterior probabilities of the BMMLR and the BMLR model were similar and did not lead to different superiority or inferiority conclusions. 
A model comparison based on Bayes factors resulted in evidence in favor of a single-level model. 
It would be helpful to have information about clustering beforehand and we concluded that these results call for a proper method to quantify the degree of dependence among observations within clusters prior to the analysis.
Such insights could help in clarifying the statistical urgency of a multilevel model and the appropriateness of a single-level model in advance.

Second, average treatment effects indicated an increased probability of recurrent events and a slightly decreased probability of long-term independent living after receiving the experimental treatment. 
However, different decision rules led to different conclusions. 
When the individual treatment effects had to be better on both dependent variables (All rule) or were weighted (Compensatory rule), no superiority or inferiority could be concluded.
When any of the dependent variables had to demonstrate a relevant treatment difference (Any rule), both inferiority on recurrent events and superiority on long-term independent living could be concluded.
This demonstrated a general potential problem with the Any rule: Contrasting decisions can result from the same analysis.
Recall that the Any rule treats all outcome variables as equally important, raising the question which conclusion to favor for patients in the Mid-High range or with a High value of severity. 
This problem does not occur with the other rules: The All and Compensatory rules are unambiguous in their conclusions.

Third, conditional (average) treatment effects suggested a trend in heterogeneity on the individual dependent variables that was not reflected by the average treatment effect. 
These trends were partially supported by superiority and/or inferiority decisions, depending on the specified decision rule. 
Even without clear conclusions, conditional treatment effect sizes provided detailed insights: Considering average treatment effects only would have overlooked these trends.
Further, the BMB model in the High range demonstrated that subgroup analysis can be a suboptimal approach to estimate conditional average treatment effects, as it can suffer from power loss. 
The High range subgroup is a relatively small fraction of the total sample size and performing an independent analysis on this group reduces the amount of evidence. 
This is reflected in the comparison to the BMMLR and BMLR methods: BMB has less extreme posterior probabilities, while treatment effect estimates are similar.

\section{Discussion}\label{s:discussion}

The current paper presented the BMMLR framework as a multilevel extension to the Bayesian multivariate logistic regression (BMLR) analysis framework. 
The BMMLR framework consisted of three elements: 
\begin{enumerate}
\item a Bayesian multilevel multivariate logistic regression model;
\item a transformation procedure to interpret results on the (multivariate) probability scale;
\item a statistical decision procedure to draw superiority and inferiority conclusions with targeted frequentist Type I errors
\end{enumerate}  
The presented framework accurately handled the multilevel structure of the data in the presence of heterogeneous treatment effects on multiple (correlated) binary dependent variables. 
A simulation study demonstrated that the proposed model indeed a) estimated average and conditional treatment effects in multilevel data without bias; and b) resulted in statistical decisions with targeted Type I error rates.
A multilevel model was clearly superior for clustered data: Naive models that did not take the multilevel structure into account resulted in inflated Type I-error rates. 
Further, the logistic model promoted information-sharing between clusters and subpopulations, being a more powerful alternative than subgroup analysis to identify heterogeneous treatment effects.
A re-analysis of the IST-3 provided another perspective on the data than the original paper \cite{IST2012}.
Detailed insights as well as the varying treatment effects among subpopulations demonstrated the importance of a) a well-considered and specific decision rule; and b) the assessment of treatment heterogeneity.  
\label{R1_1_Discussion}
The statistical need for a multilevel model has not clearly become evident for this specific analysis. 
The results suggested that a substantive cluster structure in the data does not necessarily imply a relevant statistical dependency structure between observations. 
We demonstrated that an implied dependency structure can be tested using empirical Bayes factors \cite{Vieira2023}. 
If these Bayes factors provide evidence that none of the random effects varies, a single-level model gives a more parsimonious description of the data. 
In case of evidence for the presence of random effects due to the multilevel structure in the data, the proposed multilevel multivariate model is preferred as it gives more accurate type I errors.
If there is evidence that some of the random effects do not vary across clusters, it is recommended to fix these parameters to give a more parsimonious description of the data.

Application of the BMMLR framework is not limited to the presented analyses.
Theoretically, the model can be adapted to the longitudinal setting, may be used to borrow strength from different trials, or may be extended to data with multiple levels of clustering for example. 
In practice, such extensions require additional exploration of the (computational) properties of the model, since MCMC sampling procedures appeared sensitive to the amount of autocorrelation and the number of parameters. 
In a related fashion, carefully choosing which random effects to include is helpful for smooth execution of multilevel analysis.
\label{R2_1_Discussion}
The model has a large number of options regarding specification of the model, giving a lot of flexibility to model cluster effects precisely.
This flexibility reduces parsimony however, as it easily increases the number of model parameters. 
While it is technically possible to expand the model, some care must be taken when adding many outcome variables and many covariates however.
This would result in many more model parameters, which results in considerably less parsimonious description of the data and can intensify computations notably.   
Similarly, the multinomial setup is most suitable for a limited number of dependent variables. 
Increasing the number of dependent variables results in a large number of response categories, which may lead to sparsity issues.

Future research might advance the design of the BMMLR framework in multiple ways. 
First, a priori sample size computation and power analysis have priority in medical research.
Sample sizes in logistic regression should not be too small and preferably take the success probability into account \parencite{Jong2019,Nemes2009}.
In line with our findings, larger numbers of clusters generally appear to be more powerful than larger numbers of subjects within clusters \parencite{Snijders2005}, although a study into sample sizes for multilevel logistic regression analysis provided less clear results \parencite{Moineddin2007}.
Expanding and refining knowledge regarding sample sizes in multilevel models aids in strategic experimental design \cite{Raudenbush2000, Moerbeek2000,Moerbeek2001}
Additionally, ethical aspects, such as risks and burden of (potentially inferior) treatment, and practical considerations, such as limited access to (large numbers of) subjects, require more in-depth understanding of power and sample sizes.
Especially in precision medicine – where treatments are targeted at specific patient populations - numbers of eligible subjects are limited and a priori power analysis helps to manage expectations in terms of duration.

Second, the methodology can be placed into a broader framework of Bayesian statistics.
The framework can be extended with the computation of Bayes factors to aid in decision-making regarding superiority and inferiority as well, for example following the ideas presented in Van Ravenzwaaij et al. \parencite{Ravenzwaaij2019}.
Further, the specification of prior distributions requires consideration.
Specification of non-informative priors may not be trivial.
The general tendency to choose relatively large variance parameters for normally distributed prior distributions \parencite{Gelman2008}, does not necessarily work well with the proposed model. 
Covering a range far beyond realistic parameter values, can (negatively) affect the efficiency of the sampling procedure and even the resulting posterior distribution.
Thus, concrete guidelines for the specification of non-informative priors would be helpful.

Third, pooling of treatment estimates can be done in several other ways than presented.
In general, the pooled treatment effect over clusters is a weighted combination of cluster-specific estimates, where the weights aim to balance aspects that influence estimation and are imbalanced over clusters (e.g., cluster size or variance).
Whereas we applied a cluster size-based approach, several advanced weighing procedures balance unequal variances within clusters via regularization methods \parencite[for overviews, see][]{Lin1999,Jones1998,Gallo2000}. 
These weighing methods generally produce shrinkage to the mean a) when group level variance is smaller; and/or b) when sample sizes are smaller \parencite[p. 269]{Gelman2007}.
Such weighing procedures have interesting balancing properties but are probably less suitable for trials with clusters of single subjects, such as IST-3.
These clusters have no variance, should not be discarded or merged inconsiderately, and call for the exploration of suitable weighing procedures for such data.

Finally, the BMMLR framework and multilevel models for discrete data in general lack a standard way to quantify the degree of clustering and the corresponding need for a multilevel model. 
Often, the degree of clustering is quantified as the variance between clusters relative to the variance within clusters, expressed via an intraclass correlation coefficient (ICC). 
The computation of ICCs in binary data is not straightforward: The variance within clusters - and therefore the ICC - is a function of the predictors in the model and the ICC depends on the prevalence, requiring an alternative approximation to obtain an appropriate estimate of the ICC \parencite{Ridout1999,Gulliford2005,Paul2003,Goldstein2002}. 
We leave the extension of our framework in this direction for future research.

\section{Conclusion}

The presented Bayesian method aimed to capture a multilevel structure and treatment heterogeneity simultaneously in data with multiple correlated binary outcome variables and observed covariates.
The framework was built upon three major components: a multivariate logistic regression analysis, a subsequent transformation of regression coefficients to the multivariate probability scale, and a procedure to make decisions regarding treatment superiority or inferiority.
When the sample is sufficiently large, treatment effects can be estimated unbiasedly and decisions regarding average and conditional treatment effects can be made with targeted error rates and a priori estimated sample sizes.
The method is useful in prediction of treatment effects and decision-making within subpopulations from multiple clusters, while taking advantage of the size of the entire study sample and while properly incorporating the uncertainty in a principled probabilistic manner using the full posterior distribution.

\section*{Abbreviations}
\begin{tabular}{ll}
ATE & Average Treatment Effect \\
BMB & Bayesian multivariate Bernoulli \\
BMLR & Bayesian multivariate logistic regression\\
BMMLR & Bayesian multilevel multivariate logistic regression \\
CATE & Conditional Average Treatment Effect \\
FDA & Food and Drug Administration \\
IST & Internation Stroke Trial \\
IST-3 & Third International Stroke Trial \\
NIHSS & National Institutes of Health Stroke Score \\
RCT & Randomized Controlled Trial \\
\end{tabular}


\section*{Declarations}

\subsection*{Acknowledgements}
We thank Peter Sandercock on behalf of The International Stroke Trial-3 Collaborative Group for making the data from the Third International Stroke Trial publicly available.
We gratefully acknowledge The IST-3 Collaborative Group, the trial joint sponsors (The University of Edinburgh and the Lothian Health
Board), and the chief funding agencies of the study: UK Medical Research Council, Health Foundation
UK, Stroke Association UK, Research Council of Norway, Arbetsmarknadens Partners
Forsakringsbolag (AFA) Insurances Sweden, Swedish Heart Lung Fund, The Foundation of Marianne
and Marcus Wallenberg, Polish Ministry of Science and Education, the Australian Heart Foundation,
Australian National Health and Medical Research Council (NHMRC), Swiss National ResearchFoundation, Swiss Heart Foundation, Assessorato alla Sanita, Regione dell’Umbria, Italy, and Danube University.
We thank three reviewers for their helpful comments on an earlier draft of the manuscript.

\subsection*{Funding}
The current work was supported by the Dutch Research Council (NWO) [no. 406.18.505].

\subsection*{Availability of data and materials}
The Third International Stroke Trial data that support the findings of this study are available with the identifiers [\url{https://doi.org/10.1016/S0140-6736(16)30414-7}] and [\url{http://doi.org/10.7488/ds/1350}].
The \texttt{R} code used to generate results in Sections \ref{s:evaluation} and \ref{s:application} can be found on GitHub \url{https://github.com/XynthiaKavelaars/Bayesian-multilevel-multivariate-logistic-regression}.

\subsection*{Ethics approval and consent to participate}
Not applicable

\subsection*{Competing interests}
The authors declare that they have no competing interests.

\subsection*{Author contributions}
XK performed the analyses and drafted the manuscript.
JM and MK verified analytical methods, supported the drafting of the manuscript and supervised the project.
All authors critically read and approved the manuscript.

\printbibliography

\appendix

\section{Gibbs sampling procedure based on P\'olya-Gamma expansion}\label{app:posterior_computation}

\subsection{Random effects model}
Bayesian analysis relies on the posterior distribution of regression coefficients, which is proportional to the likelihood of the data and the prior distribution: 
\begin{flalign}\label{eq:app_post}
	p(\bm{\gamma}^{q}_{j}, \bm{\gamma}^{q}, \bm{\Sigma}^{q} | \bm{y}) \propto &
	p(\bm{y}|\bm{\gamma}^{q}_{j}) p(\bm{\gamma}^{q}_{j} | \bm{\gamma}^{q}, \bm{\Sigma}^{q}) p(\bm{\gamma}^{q}) p(\bm{\Sigma}^{q}). 
\end{flalign}%

\noindent The multinomial logistic likelihood (Equation \ref{eq:H_lik_mult}) can be expanded with a P\'olya-Gamma auxiliary variable to suit a Gibbs sampling procedure. 
This expansion relies on the following equality \parencite{polson2013}:
\begin{flalign}\label{eq:H_lik_y}
	p((\bm{y}_{j} = \bm{h}^{q})|\bm{\gamma}^{q}_{j}, \bm{\gamma}^{-q}_{j}, \bm{\omega}^{q}_{j} 
	& =  \frac{\exp{(\bm{x}_{ji} \bm{\gamma}^{q}_{j})
	}}{ \displaystyle\sum_{r=1}^{Q-1} \exp{(\bm{x}_{ji} \bm{\gamma}^{r}_{j}
			)} + 1}, \\\nonumber
	& \propto \text{exp} \left[- \frac{1}{2} (
	\bm{\kappa}^{q}_{j}
	- \bm{\eta}^{q}_{j})^{T} 
	\bm{\Omega}^{q}_{j} 
	(\bm{\kappa}^{q}_{j}
	- \bm{\eta}^{q}_{j}) \right], \nonumber
\end{flalign}
\noindent where $\bm{X}_{j}$ is a matrix filled with $n_{j}$ rows of covariate vectors $\bm{x}_{ji}$ and $\bm{\eta}^{q}_{j} = \bm{X}_{j} \bm{\gamma}^{q}_{j} - \text{ln} [\displaystyle\sum_{m \neq q} \text{exp}(\bm{X}_{j} \bm{\gamma}^{m}_{j})]$, $\bm{\kappa}^{q}_{j} = \frac{I(\bm{y}_{j} = \bm{h}^{q}) - \frac{1}{2}}{\bm{\omega}^{q}_{j}}$.

Equation \ref{eq:H_lik_y} can be recognized as the kernel of a multivariate Gaussian likelihood of working variable $\bm{\kappa}^{q}_{j}$ \parencite{polson2013}:
\begin{flalign}\label{eq:H_lik_kappa}
	\bm{\kappa}^{q}_{j} 
	& \sim N \left(\bm{\eta}^{q}_{j}, \{\bm{\Omega}^{q}_{j}\}^{-1} \right) 
\end{flalign}
\noindent Here, $\bm{\Omega}^{q}_{j}$ reflects the diagonal matrix of P\'olya-Gamma distributed variables $\bm{\omega}^{q}_{j} = (\omega^{q}_{j1},\dots,\omega^{q}_{jn_{j}})$.  
A Gibbs sampler can be constructed when the likelihood in Equation \ref{eq:H_lik_kappa} is combined with multivariate normal prior distributions on random regression coefficients $\bm{\gamma}^{q}_{j}|\bm{\gamma}^{q}, \bm{\Sigma}^{q}$ and mean random regression coefficients $\bm{\gamma}^{q}$, and an inverse-Wishart prior distribution on covariance matrix $\bm{\Sigma}^{q}$:
\begin{flalign}\label{eq:app_priors}
	\bm{\gamma}^{q}_{j} & \sim N(\bm{\gamma}^{q}, \bm{\Sigma}^{q})\\\nonumber
	\bm{\gamma}^{q} & \sim N(\bm{g}^{q}, \bm{G}^{q})\\\nonumber
	\bm{\Sigma}^{q} & \sim \mathcal{W}^{-1}(j^{0}, \bm{S}^{q}) \nonumber
\end{flalign}

The resulting Gibbs sampler consists of the following steps:
\begin{enumerate}\label{list:gibbs} 
	
	\item Sample mean regression coefficients:
	$$\bm{\gamma}^{q(l)} \sim  N \left( \bm{V}^{q}_{\bm{\gamma}} (\{\bm{\Sigma}^{q(l-1)}\}^{-1} \sum_{j=1}^{J} \bm{\gamma}^{q(l-1)}_{j} + \bm{G}^{q} \bm{g}^{q}), \bm{V}^{q}_{\bm{\gamma}} 
	\right)$$
	\noindent with prior mean vector $\bm{g}^{q}$, prior precision matrix $\bm{G}^{q}$ and posterior variance matrix 
	$\bm{V}_{\bm{\gamma}} = (J \{\bm{\Sigma}^{q(l-1)}\}^{-1} + \bm{G}^{q})^{-1}$.	
	\item Sample covariance matrices of regression coefficients: 
	$$\bm{\Sigma}^{q(l)} \sim \mathcal{W}^{-1} \biggl(
	j^{0} + J, \bm{S}^{q} +  \displaystyle\sum_{j=1}^{J} \Bigl(\bm{\gamma}^{q(l)}_{j} - \bm{\gamma}^{q(l-1)} \Bigr) \Bigl(\bm{\gamma}^{q(l)}_{j} - \bm{\gamma}^{q(l-1)}\Bigr)^{T}
	\biggr)$$
	\noindent with prior hyperparameters $j^{0} \geq P$ and $\bm{S}^{q}$.
	\item For each $j$, sample random regression coefficients: 
	\begin{flalign*}
		\bm{\gamma}^{q(l)}_{j}  & \sim N \Biggl(\bm{V}^{q}_{\bm{\gamma}^{q}_{j}} \bigl(\bm{X}_{j} \bm{\Omega}^{q(l-1)}_{j} (\bm{\kappa}^{q(l-1)}_{j}
		+ \text{ln} [\displaystyle\sum_{m \neq q} \text{exp}(\bm{X}_{j} \bm{\gamma}^{m(l)}_{j})]) \bigr. \Biggr. \\\nonumber
		& \qquad{} \Biggl. \bigl. + \{\bm{\Sigma}^{q(l)}\}^{-1} \bm{\gamma}^{q(l)}\bigr), \bm{V}^{q}_{\bm{\gamma}_{j}} \Biggr)
	\end{flalign*}
	\noindent with prior mean vector $\bm{\gamma}^{q(l)}$, prior precision matrix $\bm{\Sigma}^{q(l)}$, posterior variance matrix
	$\bm{V}^{q}_{\bm{\gamma}_{j}} = ({\bm{X}_{j}}^{T} \bm{\Omega}^{q(l-1)}_{j}  \bm{X}_{j} + \{\bm{\Sigma}^{q(l)}\}^{-1})^{-1}$, and diagonal matrix of P\'olya-Gamma variables $\bm{\Omega}^{q(l-1)}_{j} = \text{diag}(\omega^{q(l-1)}_{j1},\dots,\omega^{q(l-1)}_{jn_{j}})$.
	\item For each $j$ and $i$, sample P\'olya-Gamma variables:
	$$\omega^{q(l)}_{ji} \sim PG(1, \eta^{q(l)}_{ji})$$
\end{enumerate}

\noindent The remainder of this section shows the derivations of the full conditional distributions.

\subsubsection{Deriving the likelihood function}\label{par:H_bin_lik}
The following equality forms the basis to rewrite the multinomial likelihood in Equation \ref{eq:H_lik_mult} as a Gaussian likelihood \parencite{polson2013}:
\begin{flalign}\label{eq.app_bmmlr:mult2norm}
	p((\bm{y}_{j} = \bm{h}^{q})|\bm{\gamma}_{j}, \bm{\omega}^{q}_{j}, \bm{x}_{j}) 
	& = \frac{\exp{(\bm{x}_{ji}\bm{\gamma}^{q}_{j})}}
	{ \displaystyle\sum_{r=1}^{Q-1} \exp{(\bm{x}_{ji} \bm{\gamma}^{r}_{j})
		} + 1}, \\\nonumber
	& = \prod_{i=1}^{n_{j}} 
	2 \text{ exp} \Biggl[ \kappa^{q}_{ji} \omega^{q}_{ji}
	\eta^{q}_{ji} \Biggr] \int_{0}^{\infty} \text{ exp} \Biggl[ \frac{-\omega^{q}_{ji} (\eta^{q}_{ji})^{2}}{2} \Biggr] p(\omega^{q}_{ji}) d \omega^{q}_{ji} \nonumber
\end{flalign}

\noindent where 
$\omega^{q}_{ji} \sim PG(1, \eta^{q}_{ji})$ 
is a P\'olya-Gamma distributed variable, \\
where $\eta^{q}_{ji} = \bm{x}_{ji}\bm{\gamma}^{q}_{j} - \text{ln} \Biggl[\displaystyle\sum_{m \neq q} \text{ exp}(\bm{x}_{ji}\bm{\gamma}^{m}_{j})\Biggr]$, \\
and where working variable 
$\bm{\kappa}^{q}_{j} = \frac{I(\bm{y}_{j} = \bm{h}^{q}) - \frac{1}{2}}{\bm{\omega}^{q}_{j}}$.

Further algebraic transformation results in the kernel of a Gaussian likelihood:
\begin{flalign}\label{eq.app_bmmlr:H_lik_norm}
	p((\bm{y}_{j} = \bm{h}^{q})|.)  
	& = \prod_{i=1}^{n_{j}} 
	2 \text{ exp} \Biggl[ \kappa^{q}_{ji} \omega^{q}_{ji}
	\eta^{q}_{ji} \Biggr] \int_{0}^{\infty} \text{ exp} \Biggl[ \frac{-\omega^{q}_{ji} (\eta^{q}_{ji})^{2}}{2} \Biggr] p(\omega^{q}_{ji}) d \omega^{q}_{ji}\\\nonumber
	& \propto \text{ exp} \Biggl[\frac{1}{2}( 
	\bm{\kappa}^{q}_{j} \bm{\omega}^{q}_{j}
	\bm{\eta}^{q}_{j} - \bm{\omega}^{q}_{j} (\bm{\eta}^{q}_{j})^{2} \Biggr] \\\nonumber
	& \propto \text{ exp} \Biggl[- \frac{1}{2} \left(
	\bm{\kappa}^{q}_{j}
	- \bm{\eta}^{q}_{j}\right)^{T} 
	\bm{\Omega}^{q}_{j} 
	\left(\bm{\kappa}^{q}_{j}
	- \bm{\eta}^{q}_{j}\right) \Biggr], \nonumber
\end{flalign}
Hence, working variable $\bm{\kappa}^{q}_{j}$ is multivariate normally distributed:
\begin{flalign}\label{eq.app_bmmlr:H_lik_kappa}
	\bm{\kappa}^{q}_{j} 
	& \sim N \left( \bm{\eta}^{q}_{j}, \{\bm{\Omega}^{q}_{j}\}^{-1} \right).
\end{flalign}
\subsubsection{Deriving conditional posterior distributions}

\paragraph{Random regression coefficients $\bm{\gamma}^{q}_{j}$}
Using the likelihood in Equation \ref{eq.app_bmmlr:H_lik_kappa} and prior distribution $\bm{\gamma}^{q}_{j} \sim N(\bm{\gamma}^{q}, \{\bm{\Sigma}^{q}\})$, the conditional posterior distribution of random regression coefficients $\bm{\gamma}^{q}_{j}$ is also a multivariate normal distribution:
\begin{flalign}\label{eq.app_bmmlr:H_post_random}
	p(\bm{\gamma}^{q}_{j}|.)
	\propto & \mathrlap{p(\bm{y}_{j}|\bm{\gamma}^{q}_{j}, \bm{\gamma}^{-q}_{j}, \bm{\omega}^{q}_{j}, \bm{x}) p(\bm{\gamma}^{q}_{j})} \\\nonumber
	\propto & \mathrlap{\text{ exp} \Biggl[- \frac{1}{2}
		\left(
		\bm{\kappa}^{q}_{j}
		- (\bm{\eta}^{q}_{j})
		\right)^{T} 
		\bm{\Omega}^{q}_{j} 
		\left(\bm{\kappa}^{q}_{j}
		- (\bm{\eta}^{q}_{j})
		\right) \Biggr] \times} \\\nonumber
	& \mathrlap{\text{ exp} \Biggl[ - \frac{1}{2} 
		\left(\bm{\gamma}^{q}_{j} - \bm{\gamma}^{q}
		\right)^{T} 
		\{\bm{\Sigma}^{q}\}^{-1}
		\left(\bm{\gamma}^{q}_{j} - \bm{\gamma}^{q}
		\right) \Biggr]} \\\nonumber
	\propto & \mathrlap{\text{ exp} \Biggl[- \frac{1}{2} 
		\Bigl(\{\bm{\gamma}^{q}_{j}\}^{T} (\{\bm{X}_{j}\}^{T}
		\bm{\Omega}^{q}_{j} \bm{X}_{j} + \{\bm{\Sigma}^{q}\}^{-1}) \bm{\gamma}^{q}_{j} - 2 \{\bm{\gamma}^{q}_{j}\}^{T} \Bigr.\Biggr.} \\\nonumber
	& \qquad{} \qquad{}  \Biggl. \Bigl.(\{\bm{X}_{j}\}^{T}  \bm{\Omega}^{q}_{j} (\bm{\kappa}^{q}_{j} + \text{ln} [\displaystyle\sum_{m \neq q} \text{exp}(\bm{X}_{j} \bm{\gamma}^{m}_{j})]) + \{\bm{\Sigma}^{q}\}^{-1} \bm{\gamma}^{q}) \Bigr) \Biggr] \\\nonumber
	\propto & \text{ exp} \Biggl[- \frac{1}{2} \Biggr. \\\nonumber 
	& \qquad{} 
	\Biggl(\bm{\gamma}^{q}_{j} - 
	\bm{V}^{q}_{\bm{\gamma}_{j}}
	(\{\bm{X}_{j}\}^{T} \bm{\Omega}^{q}_{j} (\bm{\kappa}^{q}_{j} + \text{ln} [\displaystyle\sum_{m \neq q} \text{exp}(\bm{X}_{j} \bm{\gamma}^{m}_{j})]) + 
	\{\bm{\Sigma}^{q}\}^{-1} \bm{\gamma}^{q})\Biggr)^{T} \\\nonumber 
	& \qquad{} \qquad{} \{\bm{V}^{q}_{\bm{\gamma}_{j}}\}^{-1} 
	\\\nonumber
	& 
	\qquad{} \Biggl. \Biggl(\bm{\gamma}^{q}_{j} -
	\bm{V}^{q}_{\bm{\gamma}_{j}}
	(\{\bm{X}_{j}\}^{T} \bm{\Omega}^{q}_{j} (\bm{\kappa}^{q}_{j} + \text{ln} [\displaystyle\sum_{m \neq q} \text{exp}(\bm{X}_{j} \bm{\gamma}^{m}_{j})]) + \{\bm{\Sigma}^{q}\}^{-1} \bm{\gamma}^{q})\Biggr)
	\Biggr] \\\nonumber
	\sim & \mathrlap{N \left(\bm{V}^{q}_{\bm{\gamma}_{j}} (\bm{X}_{j} \bm{\Omega}^{q}_{j} (\bm{\kappa}^{q}_{j}
		+ \text{ln} [\displaystyle\sum_{m \neq q} \text{exp}(\bm{X}_{j} \bm{\gamma}^{m}_{j})]) + \{\bm{\Sigma}^{q}\}^{-1} \bm{\gamma}^{q}), \bm{V}^{q}_{\bm{\gamma}_{j}} \right)}  \nonumber 
\end{flalign}

\noindent with prior mean vector $\bm{\gamma}^{q}$, prior variance matrix $\bm{\Sigma}^{q}$ and posterior variance matrix
$\bm{V}^{q}_{\bm{\gamma}_{j}} = ({\bm{X}_{j}}^{T} \bm{\Omega}^{q}_{j}  \bm{X}_{j} + \{\bm{\Sigma}^{q}\}^{-1})^{-1}$.
\paragraph{Random mean $\bm{\gamma}^{q}$}
When the posterior distribution of $\bm{\gamma}^{q}_{j}$ (Equation \ref{eq.app_bmmlr:H_post_random}) is included as a likelihood and combined with a $N(\bm{g}^{q}, \{\bm{G}^{q}\}^{-1})$ prior distribution, the conditional posterior distribution of random mean $\bm{\gamma}^{q}$ is another multivariate normal distribution:
\begin{flalign}\label{eq.app_bmmlr:H_post_gamma}
	p(\bm{\gamma}^{q}|.)
	\propto & \mathrlap{\prod_{j=1}^{J} p(\bm{\gamma}^{q}_{j}|\bm{\gamma}^{q},\bm{\Sigma}^{q}) p(\bm{\gamma}^{q})} \\\nonumber
	\propto & \prod_{j=1}^{J} \text{ exp} \Biggl[ - \frac{1}{2} 
	(\bm{\gamma}^{q}_{j} - \bm{\gamma}^{q})^{T} \{\bm{\Sigma}^{q}\}^{-1} 
	(\bm{\gamma}^{q}_{j} - \bm{\gamma}^{q}) \Biggr] 
	\times \Biggr. \\\nonumber
	& \qquad{}
	\text{ exp} \Biggl[ - \frac{1}{2} 
	(\bm{\gamma}^{q}- \bm{g}^{q})^{T} 
	\bm{G}^{q} 
	(\bm{\gamma}^{q} - \bm{g}^{q}) \Biggr] \\\nonumber
	\propto & \text{ exp} \Biggl[ - \frac{1}{2} 
	(\{\bm{\gamma}^{q}\}^{T}
	\left(J \{\bm{\Sigma}^{q}\}^{-1} \right)
	\bm{\gamma}^{q}) 
	- 2 \{\bm{\gamma}^{q}\}^{T}  
	\left(\{\bm{\Sigma}^{q}\}^{-1} \displaystyle\sum_{j=1}^{J} \bm{\gamma}^{q}_{j} \right)
	\Biggr] \times \\\nonumber
	& \text{ exp} \Biggl[ - \frac{1}{2}
	\{\bm{\gamma}^{q}\}^{T} \bm{G}^{q}
	\bm{\gamma}^{q}
	-2 \{\bm{\gamma}^{q}\}^{T} \bm{G}^{q} \bm{g}^{q}
	\Biggr] \\\nonumber
	\propto & \text{ exp} \Biggl[ - \frac{1}{2} 
	\{\bm{\gamma}^{q}\}^{T} 
	\left(J \{\bm{\Sigma}^{q}\}^{-1} + \bm{G}^{q} \right)
	\bm{\gamma}^{q} 
	- 
	\Biggr. \\\nonumber
	& \qquad{} \qquad{}
	\Biggl. 2 \{\bm{\gamma}^{q}\}^{T} 
	\left(\{\bm{\Sigma}^{q}\}^{-1} \displaystyle\sum_{j=1}^{J} \bm{\gamma}^{q}_{j} + \bm{G}^{q} \bm{g}^{q} \right)
	\Biggr] \\\nonumber
	\propto & \text{ exp} \Biggl[ - \frac{1}{2}
	\left(\bm{\gamma}^{q} - \bm{V}^{q}_{\bm{\gamma}} \left(\{\bm{\Sigma}^{q}\}^{-1} \displaystyle\sum_{j=1}^{J} \bm{\gamma}^{q}_{j} + \bm{G}^{q} \bm{g}^{q} \right)\right)^{T}
	\{\bm{V}^{q}_{\bm{\gamma}}\}^{-1} \Biggr. \\\nonumber
	& \qquad{} \qquad{} \left(\bm{\gamma}^{q} - \bm{V}^{q}_{\bm{\gamma}}
	\left(\{\bm{\Sigma}^{q}\}^{-1} \displaystyle\sum_{j=1}^{J} \bm{\gamma}^{q}_{j} + \bm{G}^{q} \bm{g}^{q} \right)
	\right)
	\Biggr] \\\nonumber
	\sim & \mathrlap{N \left( \bm{V}^{q}_{\bm{\gamma}} \left(\{\bm{\Sigma}^{q}\}^{-1} \sum_{j=1}^{J} \bm{\gamma}^{q}_{j} + \bm{G}^{q} \bm{g}^{q}\right), \bm{V}^{q}_{\bm{\gamma}} 
		\right),} \nonumber
\end{flalign}
\noindent with prior mean vector $\bm{g}^{q}$, prior precision matrix $\bm{G}^{q}$, and posterior variance matrix 
$\bm{V}_{\bm{\gamma}} = (J \{\bm{\Sigma}^{q}\}^{-1} + \bm{G}^{q})^{-1}$.	
\paragraph{Random variance $\bm{\Sigma}^{q}$}
When the posterior distribution of $\bm{\gamma}^{q}_{j}$ (Equation \ref{eq.app_bmmlr:H_post_random}) is included as a likelihood and combined with an inverse Wishart $\mathcal{W}^{-1}(j^{0}, \bm{S}^{q})$ prior, the conditional posterior distribution of random variance $\bm{\Sigma}^{q}$ is proportional to an inverse Wishart distribution:
\begin{flalign}\label{eq.app_bmmlr:H_post_var}
	p(\bm{\Sigma}^{q}|.)
	& \propto p(\bm{\gamma}^{q}_{j}|\bm{\gamma}^{q},\bm{\Sigma}^{q}) p\{\bm{\Sigma}^{q}\} \\\nonumber
	& \propto \prod_{j=1}^{J} |\bm{\Sigma}^{q}|^{\frac{1}{2}} \exp \Biggl[- \frac{1}{2}
	(\bm{\gamma}^{q}_{j} - \bm{\gamma}^{q})^{T} 
	\{\bm{\Sigma}^{q}\}^{-1} 
	(\bm{\gamma}^{q}_{j} - \bm{\gamma}^{q})
	\Biggr] \times \\\nonumber
	& \qquad{} |\bm{\Sigma}^{q}|^{\frac{1}{2} (j^{0} + p^{R} + 1)} \exp \Biggl[- \frac{1}{2} 
	tr(\bm{S}^{q} \{\bm{\Sigma}^{q}\}^{-1})\Biggr]   \\\nonumber 
	& \propto |\bm{\Sigma}^{q}|^{-\frac{1}{2}(j^{0} + J + P^{R} + 1)}  \times \\\nonumber
	& \qquad{} \exp \Biggl[- \frac{1}{2} tr \biggl( \Bigl(\bm{S}^{q} + \displaystyle\sum_{j=1}^{J} (\bm{\gamma}^{q}_{j} - \bm{\gamma}^{q}) (\bm{\gamma}^{q}_{j} - \bm{\gamma}^{q})^{T}\Bigr) 
	\{\bm{\Sigma}^{q}\}^{-1}\biggr)\Biggr]   \\\nonumber
	& \sim \mathcal{W}^{-1} \Biggl(
	j^{0} + J, \bm{S}^{q} + \displaystyle\sum_{j=1}^{J} (\bm{\gamma}^{q}_{j} - \bm{\gamma}^{q}) (\bm{\gamma}^{q}_{j} - \bm{\gamma}^{q})^{T} 
	\Biggr). \nonumber
\end{flalign}

\subsection{Mixed effects model}

A mixed effect model is defined as follows:
\begin{flalign}\label{eq:H_psi_mixed}
	\bm{\phi}^{q}_{ji} = f(\bm{x}_{ji}^{F} \bm{\beta}^{q} + \bm{x}_{ji}^{R} \bm{\gamma}^{q}_{j})
\end{flalign}
\noindent where $\bm{x}^{F}_{ji}$ and $\bm{x}^{R}_{ji}$ are vectors of fixed and random covariates respectively. 
Vectors $\bm{\beta}^{q}$ and $\bm{\gamma}^{q}_{j}$ reflect the accompanying fixed and random regression coefficients.
Function $f$ refers to the multinomial logistic likelihood function.

The multivariate normal distribution of working variable $\bm{\kappa}^{q}_{j}$ then has the following form:
\begin{flalign}\label{eq:app_H_lik_kappa_mixed}
	\bm{\kappa}^{q}_{j} 
	& \sim N \left(\bm{\eta}^{q}_{j}, \{\bm{\Omega}^{q}_{j}\}^{-1} \right).
\end{flalign}
\noindent Here, $\bm{\eta}^{q}_{j} = \bm{X}^{F}_{j} \bm{\beta}^{q} + \bm{X}^{R}_{j} \bm{\gamma}^{q}_{j} - \text{ln} [\displaystyle\sum_{m \neq q} \text{exp}(\bm{X}^{F}_{j} \bm{\beta}^{m} + \bm{X}^{R}_{j} \bm{\gamma}^{m}_{j})]$.
The likelihood in Equation \ref{eq:app_H_lik_kappa_mixed} can be combined with the prior distributions in Equation \ref{eq:app_priors},  complemented with a multivariate normally distributed prior on $\bm{\beta}^{q}$:
\begin{flalign}\label{eq:app_priors_beta}
	\bm{\beta}^{q} & \sim N(\bm{b}^{q}, \bm{B}^{q})
\end{flalign}

The Gibbs sampling algorithm in list \ref{list:gibbs} is extended with a distinct step for the fixed regression coefficients:
\begin{enumerate} 
	\item Sample fixed regression coefficients:
	\begin{flalign*}
		\bm{\beta}^{q(l)} \sim & N \left(\bm{V}^{q}_{\bm{\beta}} (\displaystyle\sum_{j=1}^{J} {\bm{X}^{F}_{j}}^{T} \bm{\Omega}^{q(l-1)}_{j}(
		\bm{\kappa}^{q(l-1)}_{j}
		- \bm{X}_{j}^{R} \bm{\gamma}^{q(l)}_{j} + \right. \\
		& \qquad{} \left. \text{ln} [\displaystyle\sum_{m \neq q} \text{exp}(\bm{X}^{F}_{j} \bm{\beta}^{m(l)} + \bm{X}^{R}_{j} \bm{\gamma}^{m(l-1)}_{j})]) + \bm{\Beta}^{q}\bm{b}^{q}), 
		\bm{V}^{q}_{\bm{\beta}} \right)
	\end{flalign*}
	\noindent with prior mean vector $\bm{b}^{q}$, prior precision matrix $\bm{\Beta}^{q}$ and posterior variance matrix
	$\bm{V}^{q}_{\bm{\beta}}  = (\displaystyle\sum_{j=1}^{J} {\bm{X}^{F}_{j}}^{T} \bm{\Omega}^{q(l-1)}_{j}  \bm{X}^{F}_{j} + \bm{\Beta}^{q})^{-1}$.
	\item Sample mean random regression coefficients:
	$$\bm{\gamma}^{q(l)} \sim  N \left( \bm{V}^{q}_{\bm{\gamma}} (\{\bm{\Sigma}^{q(l-1)}\}^{-1} \sum_{j=1}^{J} \bm{\gamma}^{q(l-1)}_{j} + \bm{G}^{q} \bm{g}^{q}), \bm{V}^{q}_{\bm{\gamma}} 
	\right)$$
	\noindent with prior mean vector $\bm{g}^{q}$, prior precision matrix $\bm{G}^{q}$ and posterior variance matrix 
	$\bm{V}_{\bm{\gamma}} = (J \{\bm{\Sigma}^{q(l-1)}\}^{-1} + \bm{G}^{q})^{-1}$.	
	\item Sample covariance matrices of random regression coefficients:
	$$\bm{\Sigma}^{q(l)} \sim \mathcal{W}^{-1} \biggl(
	j^{0} + J, \bm{\Sigma}^{0} +  \displaystyle\sum_{j=1}^{J} \Bigl(\bm{\gamma}^{q(l-1)}_{j} - \bm{\gamma}^{q(l)} \Bigr) \Bigl(\bm{\gamma}^{q(l-1)}_{j} - \bm{\gamma}^{q(l)}\Bigr)^{T}
	\biggr)$$
	\noindent with prior hyperparameters $j^{0} \geq P^{R}$ and $\bm{\Sigma}^{0}$.
	\item For each $j$, sample random regression coefficients:
	\begin{flalign*}
		\bm{\gamma}^{q(l)}_{j} \sim & N \left(\bm{V}^{q}_{\bm{\gamma}^{q}_{j}} (\bm{X}^{R}_{j} \bm{\Omega}^{q(l-1)}_{j} (\bm{\kappa}^{q(l-1)}_{j}
		- \bm{X}_{j}^{F} \bm{\beta}^{q(l)} + \right. \\
		& \qquad{} \left. \text{ln} [\displaystyle\sum_{m \neq q} \text{exp}(\bm{X}^{F}_{j} \bm{\beta}^{m(l)} + \bm{X}^{R}_{j} \bm{\gamma}^{m(l)}_{j})]) + \{\bm{\Sigma}^{q}\}^{-1} \bm{\gamma}^{q}), \bm{V}^{q}_{\bm{\gamma}_{j}} \right)
	\end{flalign*}
	\noindent with prior mean vector $\bm{\gamma}^{q(l)}$, prior precision matrix $\bm{\Sigma}^{q(l)}$ and posterior variance matrix
	$\bm{V}^{q}_{\bm{\gamma}_{j}} = ({\bm{X}^{R}_{j}}^{T} \bm{\Omega}^{q(l-1)}_{j}  \bm{X}^{R}_{j} + \{\bm{\Sigma}^{q(l)}\}^{-1})^{-1}$.
	\item For each $j$ and $i$, sample P\'olya-Gamma variables:
	$$\omega^{q(l)}_{ji}\sim PG(1, \eta^{q(l)}_{ji})$$
\end{enumerate}

\subsection{A note on prior specification}
\subsubsection{Regression parameters}
In the Gibbs sampling framework, regression coefficients are normally distributed with a mean and covariance matrix. 
We shortly discuss the role of these parameters below. 
The covariance matrix defines the spread of the distribution and therefore has a substantial influence on informativity: Small variance parameters increase prior information.
When non-informativity is preferable, large variance parameters are not the simple answer, as they may destabilize computations in Bayesian logistic regression analysis \parencite{Gelman2008}.
Jeffreys's prior could be an option, but sufficiently stable computation is not guaranteed \parencite{Poirier1994,Gelman2008}.
The challenge is therefore to specify prior variance parameters that are both sufficiently small to support stable analysis and to give a realistic support of the parameter and at the same time sufficiently large to be considered vague.

The mean hyperparameters defines the center of the distribution and becomes increasingly influential on the posterior distribution when the variance of the distribution is small. 
The relevance of adequate mean hyperparameters therefore increases with the informativity of the analysis. 
It should be noted that prior information of mean regression coefficients is not always available in the required parametrization. 
Researchers may be more likely to have information available in terms of (success) probabilities rather than logistic regression parameters.
Kavelaars et al. propose an approach to compute mean hyperparameters for the context of treatment comparison in the presence of a single patient characteristics, based on expected joint response probabilities \cite{Kavelaars2022}.

\subsubsection{Covariance matrices}
The covariance matrix follows an inverse-Wishart distribution with parameters. 
Specifying a non-informative prior on covariance matrices and variance parameters in general is not straightforward \parencite{Gelman2006, Schuurman2016}. 
The informativity of the inverse-Wishart distribution is sensitive to the size of variance parameters: small variances make inverse-Wishart distributions more informative. 
Naively specifying standard prior hyperparameters without consideration of prior information or data at hand may result in an undesirably large prior influence. 
Weakly informative (data-based) prior specification may be superior, if not essential for computational stability \parencite{Gelman2006}.

\section{Procedure for transformation to the probability scale and decision-making}\label{app:transformation}

\begin{breakablealgorithm}
	\caption{Procedure for statistical decision-making with posterior regression coefficients}\label{alg:procedure_sample}
	\begin{algorithmic}[1]
		
		\State \underline{\textbf{Step 1. Transform regression coefficients to treatment differences}}
		\State Let $\bm{\gamma}^{Q}_{j} = (0,\dots,0)$ and $\bm{x} = (1, T, w, \dots)$
		\For{draw $(l) \gets 1:L$}
		\For{cluster $j \gets 1:J$}
		
		\State \underline{\textit{Compute joint response probabilities}}

		\For{treatment $T \gets 0:1$}
		\For{joint response category $q \gets 1:Q$}
		
		\If{Population of interest defined by a range of values of $w$}\\
		\State Compute $\phi^{q(l)}_{Tj} =  \displaystyle\int_{w} \frac{ \text{exp} \left[ \bm{x}^{'}_{j} \bm{\gamma}^{q(l)}_{j}  \right] }
		{\displaystyle\sum_{r=1}^{Q-1} \text{exp} \left[ \bm{x}^{'}_{j} \bm{\gamma}^{r(l)}_{j} \right] + 1}  dw$ 
		\EndIf
		\If{Population of interest defined by a fixed value of $w$}\\
		\State Compute $\phi^{q(l)}_{Tj} = \frac{ \text{exp} \left[ \bm{x}^{'}_{j} \bm{\gamma}^{q(l)}_{j}  \right] }
		{\displaystyle\sum_{r=1}^{Q-1} \text{exp} \left[ \bm{x}^{'}_{j} \bm{\gamma}^{r(l)}_{j} \right] + 1}$
		\EndIf
		\EndFor
		
		\State \underline{\textit{Compute multivariate success probabilities}}
		\For{outcome $k \gets 1:K$}
		\State Compute $\theta^{q(l)}_{Tj} = \displaystyle\sum_{q=1}^{Q} \phi^{q(l)}_{Tj} I(\bm{h}^{q} \in \bm{U}_{k})$

		\State \underline{\textit{Compute multivariate treatment difference}}
		
		\State Compute 	$\delta^{k(l)}_{j} = \theta_{1j}^{k(l)} - \theta_{0j}^{k(l)}$
		\EndFor
		\EndFor
		\EndFor
		
		\For{outcome $k \gets 1:K$}
		\State Pool	$\delta^{k(l)} = \displaystyle\sum_{j=1}^{J} \frac{n_{j}}{\displaystyle\sum_{j=1}^{J} n_{j}} \delta^{k(l)}_{j}$
		\EndFor
		\EndFor
		\State \underline{\textbf{Step 2. Make superiority decision}}
		
		\State Define superiority region $\mathcal{S}_{R}$ 
		\State Draw conclusion
		\If{$\frac{1}{L} \displaystyle\sum_{(l)=1}^{L} I(\bm{\delta}^{(l)} \in \mathcal{S}_{R}) > p_{cut}$} Conclude superiority
		\Else{ Conclude non-superiority}
		\EndIf
	\end{algorithmic}
\end{breakablealgorithm}

\end{document}